\documentclass{aa}  
\usepackage{graphicx}
\usepackage{txfonts}
\usepackage{natbib}
\def\kms{$\rm \,km\,s^{-1}$}

\def\HH{H$_2$}
\def\msol{M$_{\odot}$}

\def\Ha{H${\alpha}$}

\def\Pa{P${\alpha}$}

\def\Pfd{Pf${\delta}$}
\def\Pfg{Pf${\gamma}$}
\def\Ba{Br${\alpha}$}

\def\Bd{Br${\delta}$}
\def\Bg{Br${\gamma}$}
\def\HeI{HeI}

\def\micro{$\rm \,\mu m$}
\def\arcsec{\,''}

\def\flux{$\rm \,erg\,s^{-1}\,cm^{-2}$}
\def\power{$\rm \,erg\,s^{-1}$}

\begin{document}
\title{Extremely massive young clusters in NGC1365}
\author{E. Galliano 
        \inst{1,2,3}
	\and
	D. Alloin 
	\inst{3,4}
	\and
	E. Pantin 
	\inst{4}
	\and
	G.L. Granato 
	\inst{5}
	\and
	P. Delva 
	\inst{3}
	\and 
	L. Silva 
	\inst{6}
	\and 
	P.O. Lagage 
	\inst{4}
	\and
	P. Panuzzo 
	\inst{4,5}
        }
\institute{Observat\'orio Nacional, Rua General Jos\'e Cristino, 77, 20921-400, S\~ao Cristov\~ao, Rio de Janeiro, Brazil\\
           \email{egallian@on.br}
	   \and 
	   Departamento de Astronomia, Universidad de Chile, Casilla 36-D, Santiago, Chile
           \and
           European Southern Observatory, Casilla 19001, Santiago 19, Chile
	   \and 
	   Laboratoire AIM, CEA/DSM-CNRS-Universit\'e Paris Diderot, IRFU/Service d'Astrophysique, B\^at.709, CEA/Saclay, F-91191 Gif-sur-Yvette Cedex, France
           \and 
	   INAF - Osservatorio Astronomico di Padova, Vicolo Osservatorio 5, 35122 Padova, Italy	  
	   \and
	   INAF - Osservatorio Astronomico di Trieste, Via Tiepolo 11, 34131 Trieste, Italy
          }


          \date{Received --- ; accepted --- } 
          
          \abstract {In a previous work, three bright mid-infrared/radio sources were discovered in the nuclear region of
            startburst/AGN galaxy NGC1365.} {The present study aims at
            confirming that these sources are indeed young and massive
            ``embedded'' clusters, and at deriving their physical
            parameters, such as extinction, age and mass.}  {Using
            ISAAC and VISIR at the VLT we obtained maps and low
            resolution spectra in the near- and mid-infrared. The
            resulting datasets are first interpreted by comparing the
            observations with images and spectra of the close-by young
            cluster R136 in the Large Magellanic Cloud and then by
            using model predictions for both the nebular emission
            lines and the spectral energy distribution of the
            sources.}  {We produce maps of the region containing the
            three sources in the R, J, Ks, L' bands and at
            12.8\micro~and perform their accurate relative
            positioning. We also provide spectra in the ranges
            1.8-2.4\micro, 3.3-4.0\micro, 8.1-9.3\micro~and
            10.4-13.2\micro. The spectral energy distribution of the
            three sources rises with wavelength.  Emission lines from
            ionised hydrogen and molecular hydrogen are detected, as
            well as PAH emission.  Conspicuous [NeII]12.8\micro~line
            is also present, while neither the [ArIII]\,8.9\micro~nor
            the [SIV]\,10.4\micro~lines are detected.  This provides a
            stringent constraint on the age of the sources: we argue
            that they are relatively evolved young clusters
            (6-8\,Myr). Owing to their ionising photon emission rates
            and ages, they must be extremely massive clusters (of the
            order of $10^7$\,\msol). Their mid-infrared spectral
            energy distribution suggests the presence of two
            components: (1) an optically thin component, with a
            continuum comparable to that of R136, and (2) an optically
            thick component which might be related to subsequent or
            on-going episodes of star formation. We anticipate that
            these sources are good candidates for evolving according
            to a bi-modal hydrodynamical regime, in which matter is
            trapped at the centre of a compact and massive cluster and
            generates further star formation.}  {}

          
          \authorrunning{Galliano et al.}  

          \titlerunning{Extremely massive star clusters in NGC1365}
          
          \maketitle

\section{Introduction}
\label{Introduction}
Starburst regions in close-by galaxies were first resolved in a
population of star clusters in the early nineties, thanks to the high
angular resolution of the Hubble Space Telescope. The first
galaxy-target, NGC1275 \citep{Holtzman92}, showed a population of
young and massive compact clusters.  Soon after, similar objects were
encountered in a wide variety of environments, such as dwarf galaxies
\citep{O'Connell94,Hunter94,O'Connell95,Leitherer96,Gorjian96},
interacting galaxies
\citep{Whitmore93,Conti94,Shaya94,Whitmore95,Meurer95} and
circumnuclear star-forming rings
\citep{Benedict93,Barth95,Bower95,Maoz96}.  Their possible parental
link with classical and well-studied globular clusters was proposed
and, in subsequent studies, these young massive clusters (abbreviated
YMCs) became increasingly referred to as objects likely to evolve into
globular clusters after a few Gyrs. Their masses are greater than
$10^5$\,\msol, with radii smaller than 5\,pc and ages below
100\,Myr. In the literature, the term super star cluster (SSC) is
often associated with YMCs which are bright in the visible, hence
suffer little extinction. We prefer to use the generic term YMC in all
cases, and specify, whenever needed, whether the YMC is still dust
embedded (embedded YMC) or naked (UV-bright YMC).

The youngest YMCs discovered so far may be younger than 1\,Myr, and
show up as heavily dust-embedded HII regions. Examples of such
extragalactic embedded YMCs known to date are still scarce: some were
found and discussed in the Antennae galaxies NGC4038/39
\citep{Mirabel98,Gilbert00}, in Henize 2-10 \citep{Kobulnicky99}, in
NGC5253 \citep{Gorjian01}, in SBS 0335-052 \citep{Plante02}, in IIZw40
\citep{Beck02}, in NGC1365 and NGC1808 \citep{Galliano05a,Galliano08a}, in
NGC7582 \citep{Wold06}.

As proposed in \citet{Johnson04}, parallel evolutionary sequences can
be imagined for the formation of massive stars and for the formation
of massive clusters. Both types of objects start as HII regions deeply
embedded in a dust cocoon. In the case of a massive star, this stage
is identified as ultra compact HII region (UCHII region) and in the
case of a YMC as ultra dense HII region (UDHII region)
\citep{Kobulnicky99}. They are both inconspicuous in the visible and
near-infrared (NIR), while bright in the mid-infrared (MIR) and
far-infrared (FIR). They are also intense sources of thermal radio
continuum as well as of line emission from ionised gas. The embedding
material eventually dissipates off and they become detectable as
UV-bright sources. The extent to which this parallel is sustainable
remains an open question: similarities and differences along the two
sequences are worth investigating and might bring clues about the
conditions for star formation in galaxies, and in particular for the
formation of massive star clusters and globular clusters.

We present in this paper the first step of a thorough analysis and
modelling of embedded YMCs in nearby spiral galaxies. The three YMCs
we are interested in here are located in the starburst circumnuclear
region of the Seyfert2 galaxy NGC1365 (distance 18.6\,Mpc, hence
1\arcsec~corresponds to 90\,pc). \citet{Galliano05a} discovered these
sources in the MIR and demonstrated that they coincide with bright
thermal centimetre radio sources detected by \citet{Sandqvist95}. They
have also been recently detected in the CO molecule by
\citet{Sakamoto07}. These authors infer a mass of molecular material
of the order of 10$^9$\msol~in the central 2\,kpc diameter region,
which includes the three YMCs under discussion.

In \citet{Galliano05a}, their ages were grossly estimated from their
radio spectral indexes and found to be of a few (~3) Myrs. Accordingly,
their masses were estimated to be of the order of 10$^6$\,\msol.

These clusters are located at the inner Linblad resonance (ILR) in
their host galaxy, and also within around~1\,kpc from its active
galactic nucleus (AGN). Their environment has a slightly above solar
metallicity, as expected from the observed metallicity gradient in the
disc of NGC1365 discussed by e.g. \citet{Dors05}.  

The questions to be addressed in the current paper remain basic
ones. What are the spectral characteristics of such YMCs? Are NIR
and/or MIR data sufficient to get a trustable insight on the
properties of embedded YMCs? Do we miss a substantial part of the
phenomenon by observing only in the NIR? Which physical parameters can
be safely derived for these objects, either in a direct way or through
a comparison with models?

The paper is structured as follows: the first part is devoted to the
presentation and discussion of the data, while subsequent sections
deal with the physical interpretation of the sources. In
Sec.~\ref{Data}, we describe the acquisition and reduction of the
dataset, which consists of NIR/MIR images and spectra, all collected
at the ESO/VLT using the instruments ISAAC and VISIR. We first discuss
the images and perform the relative registration of the maps at
different wavelengths (Sec.~\ref{Registration}). We then describe and
discuss the spectra (Sec.~\ref{Spectra}), which display intense
nebular lines and NIR/MIR rising continua. In Sec.~\ref{Measurements},
we discuss the uncertainties on the quantities measured from the
images and from the spectra.

In Sec.~\ref{R136}, we perform a comparison of the data for the three
YMCs in NGC1365 with comparable data obtained for R136 in the LMC, one
of the nearest known YMC.  To do so, we use WFI and IRAC images, an
ISO SWS spectrum and a wide field ISO CVF spectrum of R136. In
Sec.~\ref{YMC basic parameters}, we derive basic parameters for the
YMCs based upon their emission lines, via a comparison with
predictions from a library of photo-ionisation models generated with
the code \texttt{CLOUDY}.  In Sec.~\ref{SED}, we analyse the NIR/MIR
infrared emission of the sources, both their spectral energy
distribution (SED) and their line emission, performing a more complex
modelling with the dusty stellar population evolution code
\texttt{GRASIL}.  Then, in Sec~\ref{specific evolution}, we attempt to
position our results in the light of the theoretical evolution of very
massive clusters, considering the bi-modal hydrodynamic solution for
re-inserted matter, as proposed by \citet{Silich07}~and
\citet{Tenorio07}. Finally, the conclusions and perspectives of our
work are highlighted in Sec.~\ref{conclusion}.


\section{The NIR/MIR dataset}
\label{Observations}

Using the ESO infrared facilities, we obtained new images of the
central region of NGC1365, as well as spectra of the three MIR/radio
sources unveiled by \citet{Galliano05a}. With ISAAC, the NIR
spectro-imager at VLT/UT1, we have collected J
(1.2\micro)\footnote{P074.B-0166; October 29-30, 2004}, Ks
(2.2\micro), L' (3.8\micro) and M (4.5\micro)\footnote{P072.B-0397;
  December 01-02, 2003} images at an angular resolution of the order
of 0.6\arcsec, and low resolution long slit spectra of the three
MIR/radio sources, in the K and L bands$^2$. Notice that throughout
this paper, we retain the source nomenclature as in
\citet{Galliano05a}: the embedded sources are referred to as M4, M5
and M6.  For one of the sources (M6), we also obtained a spectrum in
the N band (around 10\micro)\footnote{P074.B-0166; October 23-24,
  2004} with TIMMI2, the MIR spectro-imager of the 3.6m telescope at
La Silla: despite its rather low S/N ratio in the continuum, a
prominent [NeII] line could be detected. This encouraged us to perform
additional observations with VISIR at VLT/UT3, providing an image in
the narrow [NeII] filter at 12.8\micro\footnote{P074.A-9016; December
  01, 2004}, together with low resolution spectra in the 8\micro,
11\micro~and 12\micro~bands \footnote{P076.B-0374(A); November 18 \&
  20, 2005}.

In addition, we use in our analysis a WFPC2 R band image retrieved
from the HST archive, the ATCA centimetre maps by \citet{Forbes98} and
by \citet{Morganti99}, as well as the centimetre measurements of
\citet{Sandqvist95}.  We also consider the TIMMI2 10.4\micro,
11.9\micro~and 12.9\micro~images previously obtained by
\citet{Galliano05a}.

\subsection{Data collection and reduction}
\label{Data}

All data were acquired and reduced using standard techniques. Let us
briefly recall the main steps below:

\textbf{ISAAC images}: The images (J, Ks, L' and M bands) were
obtained using the Aladin detector, with pixel scales of
0.148\arcsec~per pixel in the J and Ks bands and 0.071\arcsec~per
pixel in the L' and M bands. The on-source integration times for the
J, Ks, L' and M bands were respectively of 90\,sec, 300\,sec,
300\,sec, 450\,sec. A standard nodding technique was applied for the J
and Ks observations, while chopping and nodding were used for the L'
and M observations. All images were reduced with the \texttt{ECLIPSE}
package. The photometric calibration of the images relied on
observations of a standard star. The precision on photometric
measurements is of the order of 10$\%$. The achieved angular
resolutions are 0.56\arcsec~in J, 0.44\arcsec~in Ks, 0.39\arcsec~in L'
and 0.38\arcsec~in M. The M band image has a low S/N ratio though and
is not used in the following.

\textbf{ISAAC spectra}: The low resolution long slit mode was used to
collect Ks and L' band spectra of the sources M4, M5 and M6. Spectral
resolutions were respectively R=450 and 360 for the Ks and L'
bands. Two positions of the 1\arcsec~width slit allowed to obtain the
three spectra: one slit at PA=2.8\degr~(PA positive from North to
East) passed through the AGN and M4, while another slit at
PA=145.1\degr~ passed through M5 and M6. The slits were precisely
positioned following the measurements in \citet{Galliano05a}, and by
performing blind offsets referenced on the AGN. Standard data
reduction procedures were applied, using \texttt{ECLIPSE} and
\texttt{IRAF}. In the Ks band, self-chopping effects prevent assessing
precisely the continuum level, although they do not affect the
measurement of emission line fluxes. The spectra were extracted
through slit windows of 1.4\arcsec~along the slits.  Night-sky lines
were used for the wavelength calibration.

\textbf{TIMMI2 spectrum}: With TIMMI2, an N band spectrum was obtained
for M6, using the 10\micro~low resolution grism (7.5\micro~to
13.9\micro) with spectral resolution R=160. The pixel scale is
0.45\arcsec. The slit, 1.2\arcsec~wide and at PA= 0\degr, was blindly
positioned on the target using the AGN as a reference. The standard
nodding and chopping technique allowed an efficient removal of the
background emission. The on-source exposure time was about
4000\,sec. Subsequent data processing consists in the addition of
chopping and nodding pairs, followed by a shift-and-add procedure to
sum the two negative and one positive spectra generated by the
chopping-nodding technique.

\textbf{VISIR image}: VISIR allowed to collect an image of NGC1365
through the narrow [NeII] 12.8\micro~filter (FWHM of 0.2\micro). The
standard chopping and nodding technique enabled to remove the MIR
background. The data reduction consists in shifting and stacking the
individual frames (each corresponding to a chopping position). The
achieved angular resolution is 0.4\arcsec. To optimise the detection
of extended features, the image was filtered using the
\texttt{mr\_filter} routine of the \texttt{MR/1} software package
developed by \citet{Murtagh99}.

\textbf{VISIR spectra}: We chose for the VISIR spectra the same two
slit positions as for the ISAAC spectra. We observed in three settings
with centred wavelengths at 8.5\micro, 9.8\micro~and 12.2\micro~, in a
low resolution mode ($\lambda/\Delta\lambda$ of the order of
200). Exposure times were respectively 655\,sec, 510\,sec and 516\,sec
for the three settings. Doing so, we achieve the following spectral
coverage: from 8.1\micro~to 9.3\micro, and from 10.4\micro~to
13.0\micro, with a good overlap around 11\micro. Observing procedures
and data reduction techniques are similar to those described for the
TIMMI2 spectra. The absolute calibration of the 12.2\micro~centred
spectrum was assessed through the calibrated VISIR 12.8\micro~narrow
band image. The 9.8\micro~centred spectrum was then scaled, through
its overlap with the 12.2\micro~centred spectrum. For the spectrum
centred at 8.5\micro~, we retain the original flux
calibration. Therefore, there is some uncertainty between the relative
fluxes of the 8.1-9.3\micro~and the 10.4-13\micro~segments of the
spectrum. The precision on the absolute calibration of the 10.4 to
13\micro~segment is of the order of 20$\%$, comparable to that of the
VISIR image and better than that of the 8.1-9.3\micro~segment. The
calibration uncertainty for the latter spectral segment is mostly due
to possible varying slit losses, and might reach 40$\%$.

\begin{figure}[htbp]
\begin{center}
\includegraphics[width=8.5cm]{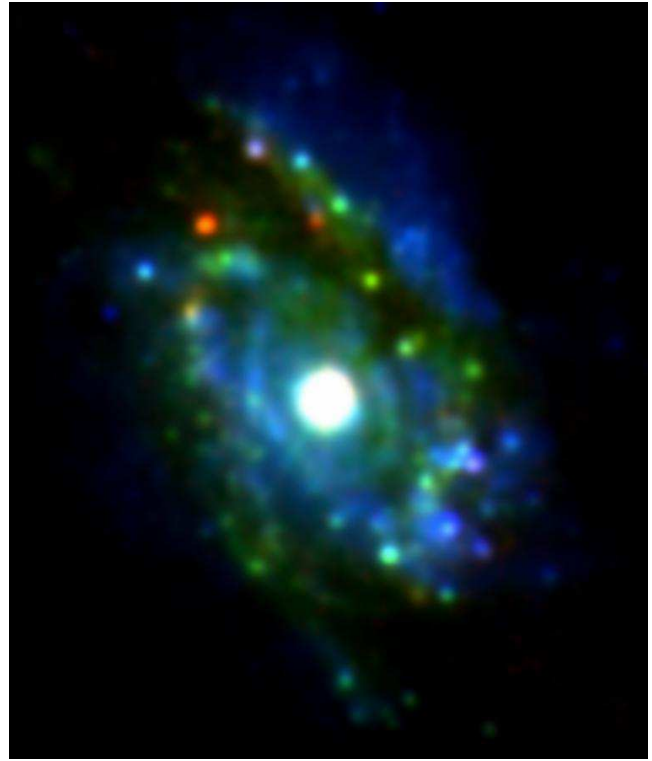}
\caption{Three colour image of the central region of NGC1365. The WFPC2 R band image is blue, the ISAAC Ks band image is green, and the ISAAC L' band image is red. North is up and East is to the left. The dimensions of the 
image, 25\arcsec$\times$30\arcsec correspond to 2.25\,kpc$\times$2.7\,kpc~for a distance of 18.6\,Mpc} 
\label{fig colour}
\end{center}
\end{figure} 
\begin{figure*}[htbp]
\begin{center}
\includegraphics[width=18cm]{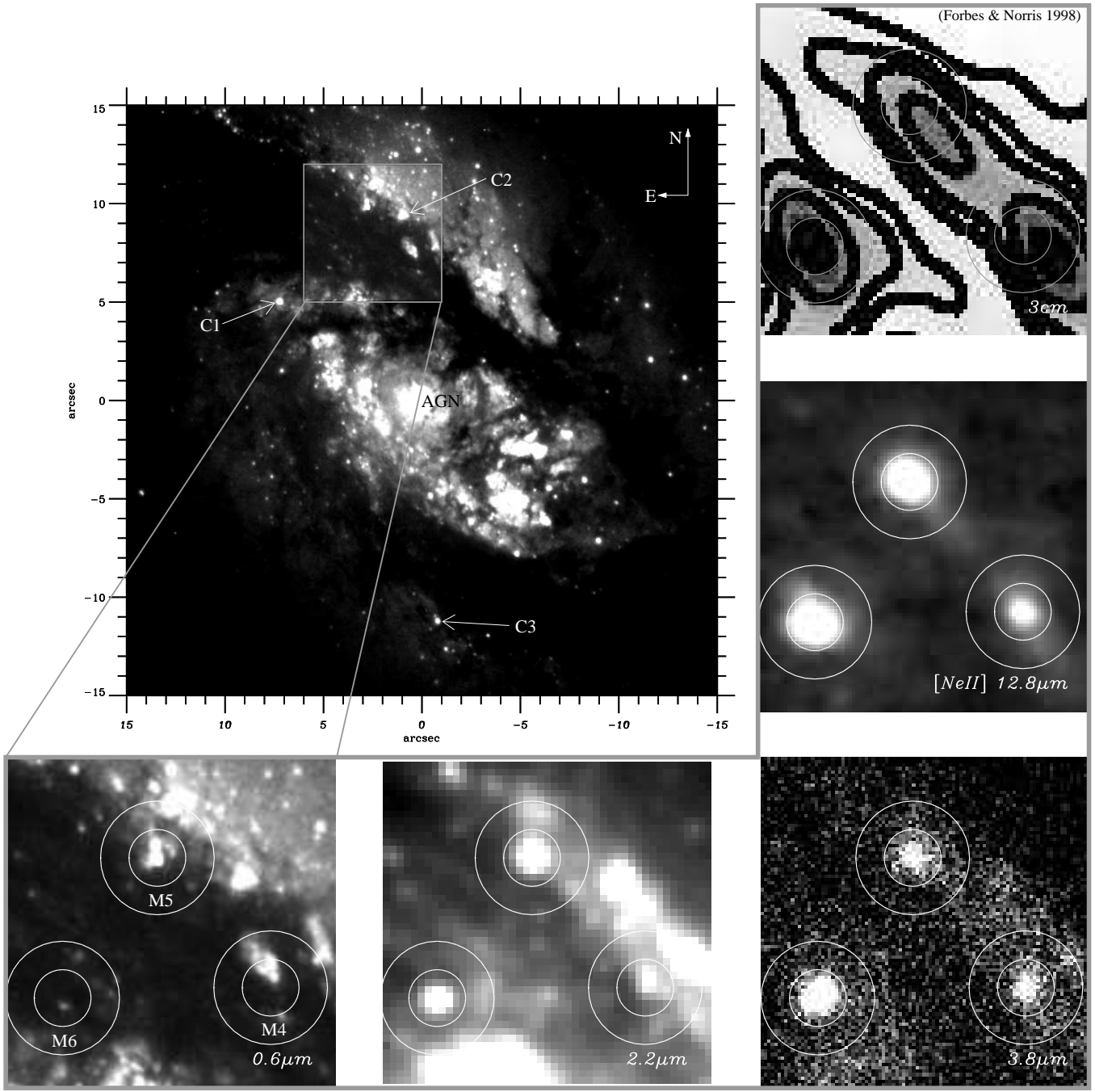}
\caption{The embedded clusters in NGC1365. The large image shows the
  inner 30\arcsec $\times$ 30\arcsec~observed with HST/WFPC2 in the
  F606W filter (0.6\micro). The highlighted square shows the region
  where the three embedded young massive star clusters M4, M5 and M6
  are located. The small images detail this region at five
  wavelengths: 0.6\micro~(HST/WFPC2), 2.2\micro~(VLT/ISAAC),
  3.8\micro~(VLT/ISAAC), 12.8\micro~(VLT/VISIR) and 3\,cm
  \citep[][ ATCA, reproduced from their work]{Forbes98}. For clarity,
  the J (1.2\micro) image mentioned in the text is not displayed
  here. Its angular resolution is rather low and adding it would not
  bring any pertinent information. The locations of M4, M5 and M6 are
  given by the centres of the concentric circles on the small
  images. They correspond to the locations of the
  12.8\micro~sources. Sources labelled C1, C2 and C3 as well as the
  active galactic nucleus (labelled AGN) are used for the registration
  (see Sect.~\ref{Registration}).}
\label{fig summary}
\end{center}
\end{figure*} 
\begin{figure*}[htbp]
\begin{center}
\includegraphics[width=14cm]{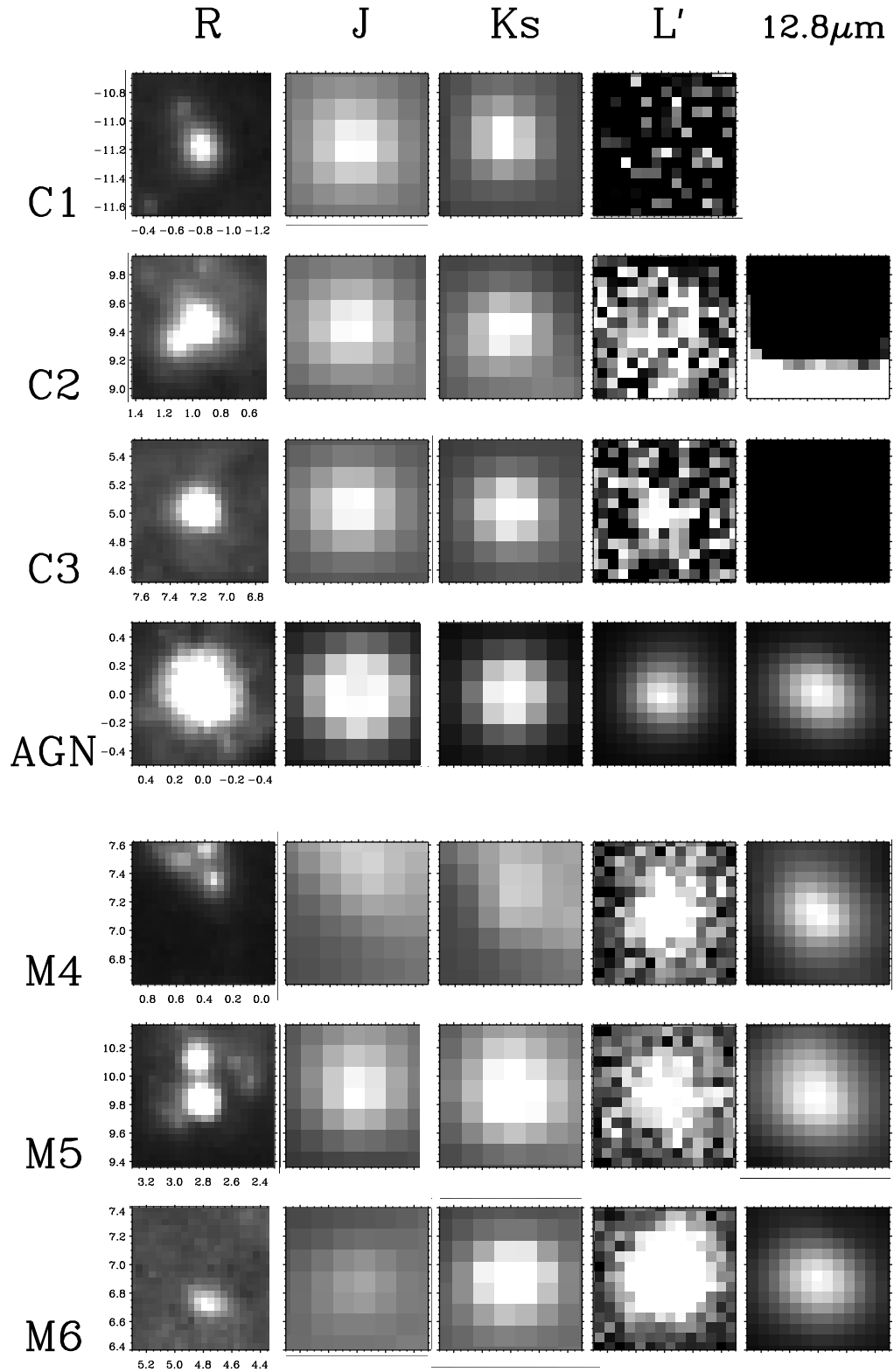}
\caption{Details of the different images illustrating, for C1, C2 and
  C3 the quality of the registration, and for M4, M5 and M6, the
  morphological changes that these sources suffer with
  wavelength. Each raw shows the images of a given source at the
  various observed wavelengths. The wavelengths are from left to
  right: R, J, Ks, L' and 12.8\micro. The sources are from top to
  bottom: C1, C2, C3 (see Fig.~\ref{fig summary}), AGN, M4, M5,
  M6. The first four raws illustrate the quality of the
  registration. The last three raws highlight the morphological
  changes of the embedded YMCs with wavelength. The X and Y scales for
  each image represent the same coordinates as on Fig.~\ref{fig
    summary}, where (0\arcsec;\,0\arcsec) coincides with the AGN
  location.}
\label{fig registration}
\end{center}
\end{figure*} 
\begin{figure}[htbp]
\begin{center}
\includegraphics[width=8cm]{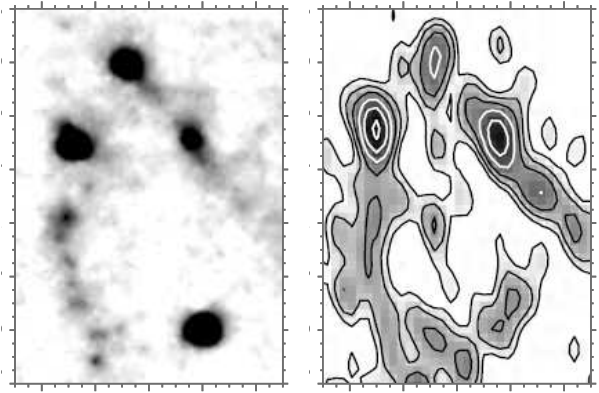}
\caption{Comparison between the VISIR narrow band 12.8\micro~filter image (left) and the 3.5cm map (right) by \citet{Morganti99} for a 
10\arcsec $\times$ 14\arcsec region including the AGN (bottom source on the left image), the three embedded YMCs and the MIR/cm ring. North is up, East is to the left. The ticks along the axis are drawn with 0.5\arcsec~intervals.}
\label{fig MIR-cm}
\end{center}
\end{figure} 

\begin{figure*}[htbp]
\begin{center}
\includegraphics[width=18cm]{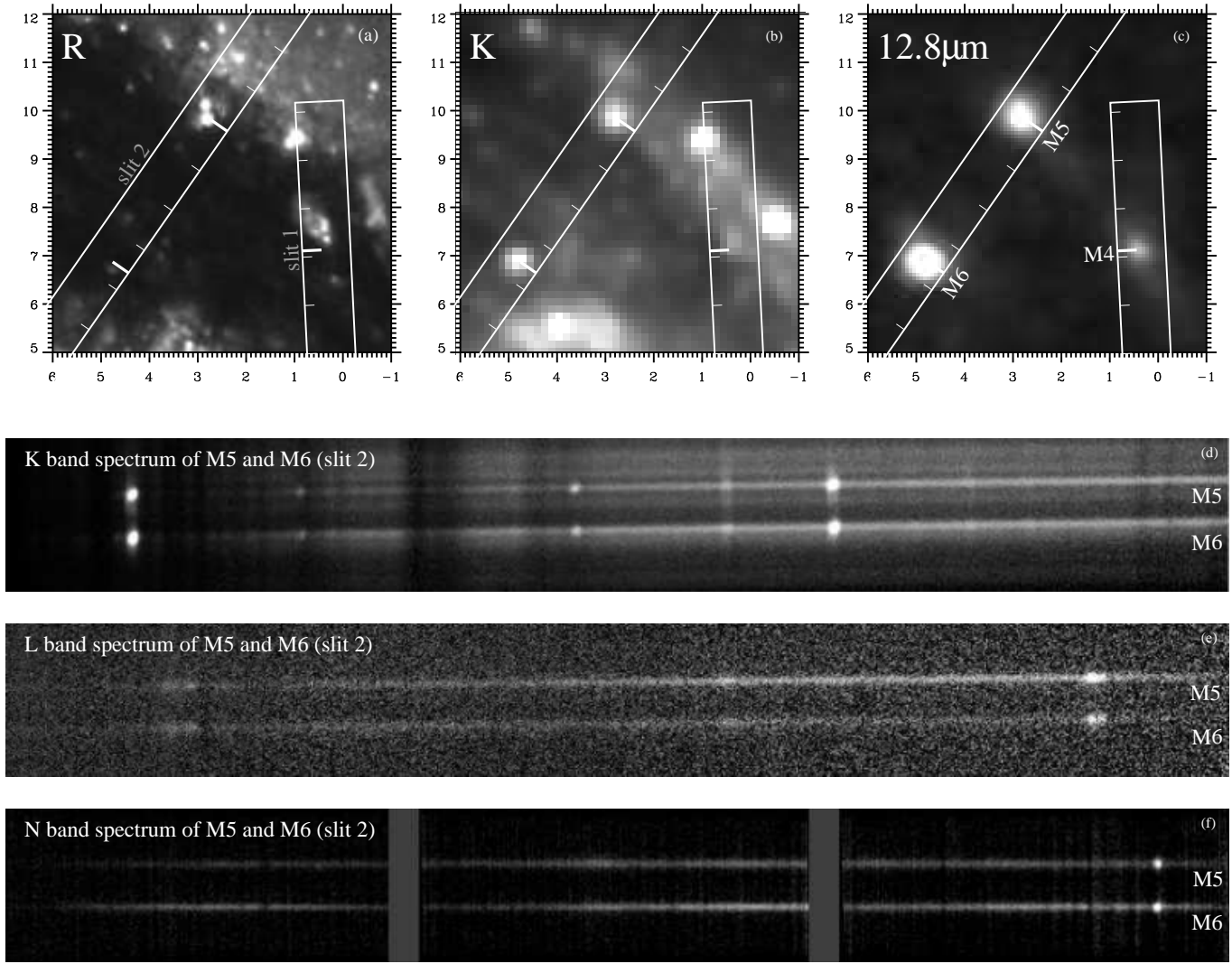}
\caption {\textbf{Frames (a, b, c)} from left to right: R, Ks and 12.8\micro~images of the embedded clusters region. North is up and East is to the left. The slit drawings show the positions of slit 1 (through the nucleus and M4) and slit 2 (through M5 and M6). Along the slits, the thin and short ticks mark a scale in arcsec, while the thick and long ticks mark the position where the spectra are recorded. The references are the nucleus for slit 1, and M5 for slit 2. \textbf{Frames (d \& e)}: ISAAC spectra through slit 2 after reduction and before extraction. The wavelength increases from left to right. In the Ks spectrum, the two brightest lines are \Pa~and \Bg. This image shows that, due to self-chopping, it is difficult to assess the actual continuum level. In L', the bright line to the right is \Ba. Since, no extra emission is detected along the slit --apart from the two clusters-- self chopping is inexistent. \textbf{Frame (f)} VISIR spectra, from left to right separated by grey bands: 8.5\micro~setting (8.1-9.3\micro), 9.8\micro~setting (10.4-12.4\micro), 12.2,\micro~setting (11.3-13.0\micro). The bright emission line on the right is [NeII]. }
\label{fig spectra}
\end{center}
\end{figure*} 
\begin{figure}[htbp]
\begin{center}
\includegraphics[width=8cm]{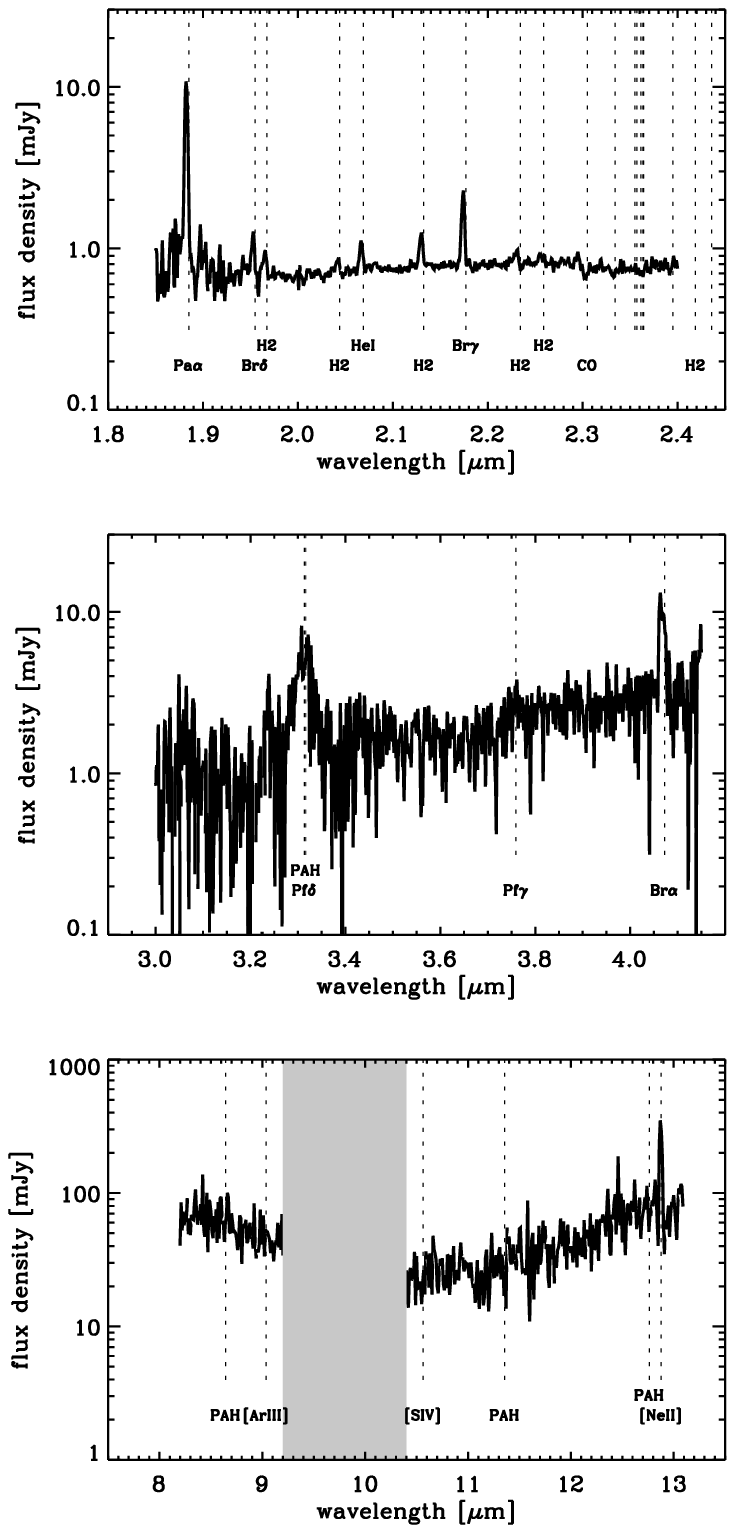}
\caption{From top to bottom, Ks, L' and N band spectra of M4.}
\label{fig spectrum M4}
\end{center}
\end{figure} 
\begin{figure}[htbp]
\begin{center}
\includegraphics[width=8cm]{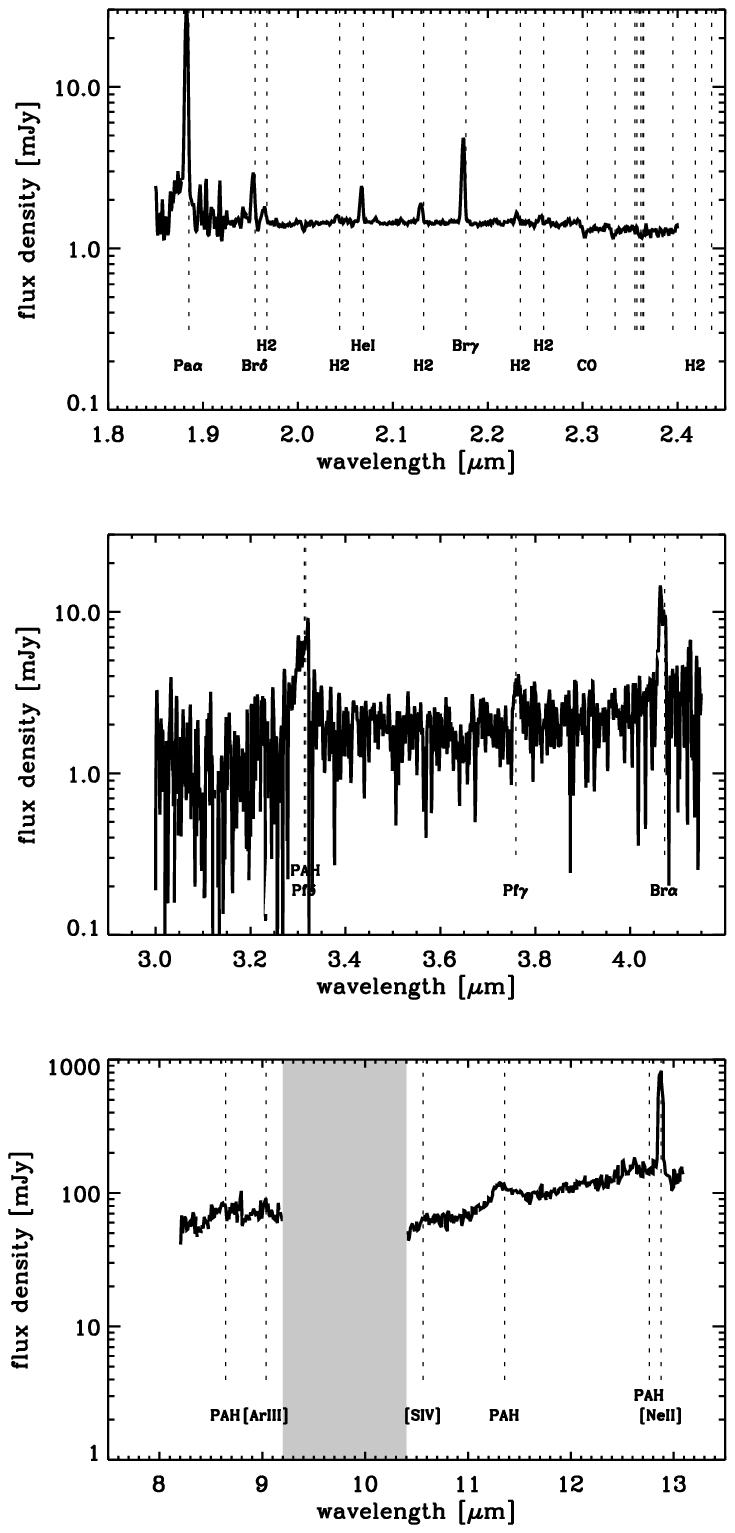}
\caption{From top to bottom, Ks, L' and N band spectra of M5.}
\label{fig spectrum M5}
\end{center}
\end{figure} 
\begin{figure}[htbp]
\begin{center}
\includegraphics[width=8cm]{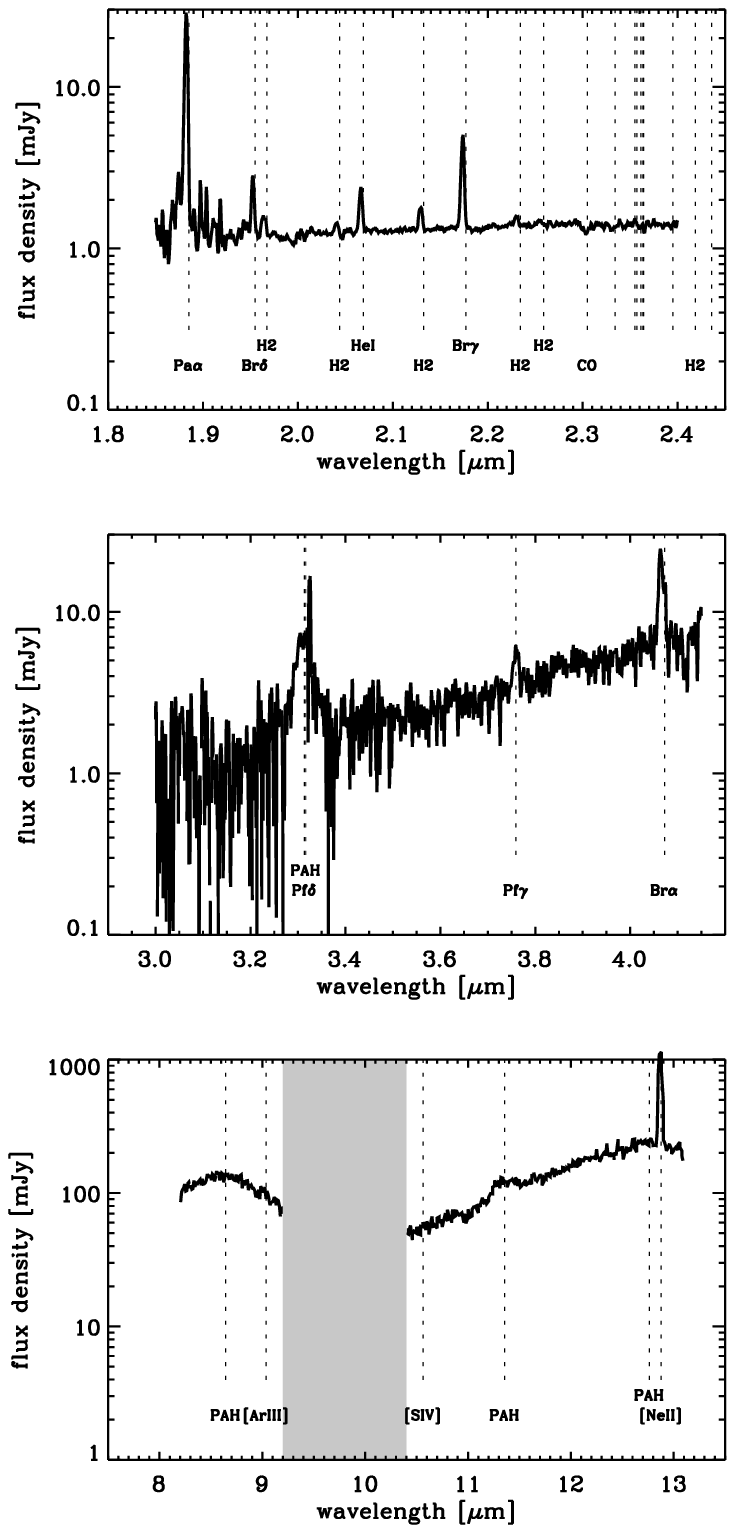}
\caption{From top to bottom, Ks, L' and N band spectra of M6.}
\label{fig spectrum M6}
\end{center}
\end{figure} 
\begin{figure}[htbp]
\begin{center}
\includegraphics[width=8cm]{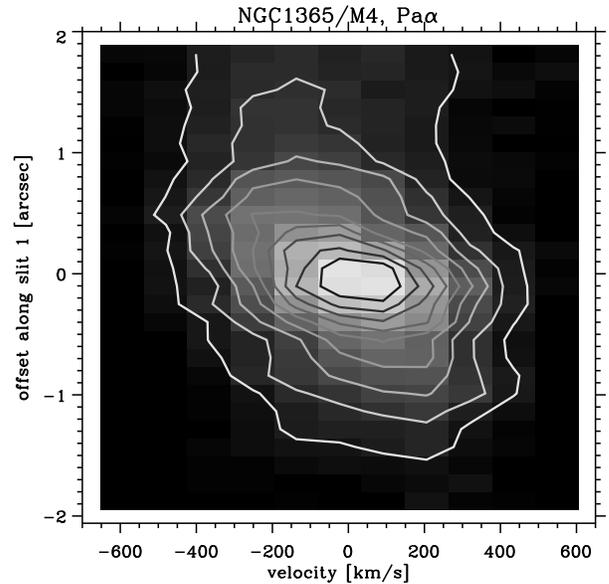}
\caption{Position-Velocity (\Pa~line) diagram for M4 along slit 1 (PA=-2.8\degr). The X-axis represents the velocity, the X-0 value being set at the emission peak. The Y-axis represents the spatial direction, top corresponding to the North. The Y-0 coordinate corresponds to the emission peak, also marked by a thick tick along the drawing of slit 1 in Fig.~\ref{fig spectra}. At any position, the instrumental broadening is dominant (660\,\kms).}
\label{fig PV}
\end{center}
\end{figure} 
\begin{figure}[htbp]
\begin{center}
\includegraphics[width=8cm]{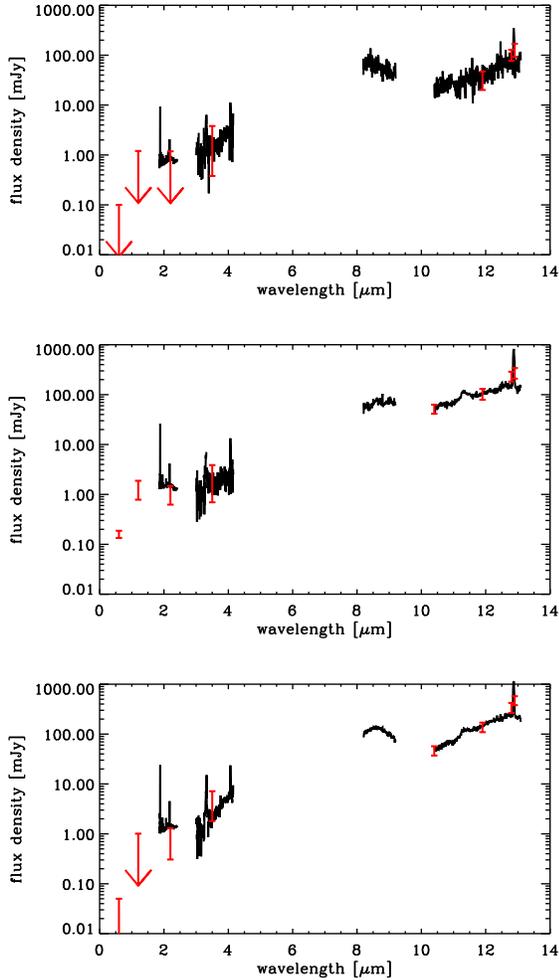}
\caption{NIR/MIR SEDs of the embedded clusters M4, M5 and M6 in NGC1365 (red points with error bars or upper limits), over-plotted on the Ks, L' and N band spectra (black continuous line). The NIR measurements are from this article, while the MIR points are from \citet{Galliano05a}. For M4, since no direct Ks band counterpart is observed on the image, the photometric point is given as an upper limit. This apparent inconsistency denotes the fact that we do not formally associate the Ks band continuum to the cluster emission (see text).} 
\label{fig SED}
\end{center}
\end{figure} 

\begin{table*}[htbp]
\begin{center}
\caption[]{Measurements derived from the spectra.}
\label{table measurements}
\begin{tabular}{lcccccc}

\hline\hline
                                                                                   & \multicolumn{2}{c}{M4}   & \multicolumn{2}{c}{M5}   & \multicolumn{2}{c}{M6} \\ 
\hline
                                                                                   & \multicolumn{2}{c}{value}& \multicolumn{2}{c}{value}& \multicolumn{2}{c}{value} \\ 
                                                                                   & low        & high       & low        & high       & low        & high      \\ 
\hline
\multicolumn{7}{c}{position with respect to AGN [arcsec]}\\
                                                            $\Delta\alpha$~[arcsec]&       0.31 &       0.51 &       2.70 &       2.90 &       4.70 &       4.90 \\ 
                                                            $\Delta\delta$~[arcsec]&       7.02 &       7.22 &       9.76 &       9.96 &       6.80 &       7.00 \\
\hline 
\multicolumn{7}{c}{flux densities [mJy]}\\
                                                                                R  &\multicolumn{2}{c}{$\le$0.10} &       0.13 &       0.19 & \multicolumn{2}{c}{$\le$0.05} \\ 
                                                                                J  &\multicolumn{2}{c}{$\le$1.21} &       0.79 &       1.88 & \multicolumn{2}{c}{$\le$1.01} \\ 
                                                                                Ks &\multicolumn{2}{c}{$\le$1.18} &       0.62 &       1.47 &       0.31 &       1.29 \\ 
                                                                                L' &       0.28 &       3.80 &       0.69 &       3.87 &       1.81 &       7.16 \\ 
                                                                      10.4\micro~  &       0.57 &       3.94 &      41.33 &      62.90 &      37.18 &      57.01 \\ 
                                                                      11.9\micro~  &      20.10 &      47.26 &      78.76 &     130.16 &     109.15 &     169.90 \\ 
                                                                      12.9\micro~  &      91.97 &     169.03 &     208.83 &     341.28 &     379.61 &     575.43 \\ 
                                                                      12.8\micro~  &      77.24 &     128.32 &     183.10 &     287.64 &     264.64 &     421.92 \\ 
                                                                              6cm  &       2.38 &       2.98 &       1.12 &       1.72 &       2.45 &       3.05 \\ 
                                                                             20cm  &       3.04 &       3.64 &       2.44 &       3.04 &       2.24 &       2.84 \\ 
\hline
\multicolumn{7}{c}{line fluxes [10$^{-15}$\,erg\,s$^{-1}$\,cm$^{2}$]}\\
                                           \Pa                                     &      23.18 &      34.76 &      64.64 &      96.96 &      60.43 &      90.65 \\ 
                                           \Bd                                     &       1.23 &       1.84 &       3.40 &       5.10 &       3.48 &       5.22 \\ 
                                   \HH~1-0S(3)                                     &       0.71 &       1.07 &       0.96 &       1.44 &       1.29 &       1.94 \\ 
                                   \HH~1-0S(2)                                     &       0.36 &       0.54 &       0.37 &       0.56 &       0.44 &       0.66 \\ 
                                           HeI                                     &       0.86 &       1.29 &       1.85 &       2.77 &       2.44 &       3.67 \\ 
                                   \HH~1-0S(1)                                     &       1.06 &       1.59 &       1.02 &       1.53 &       1.13 &       1.70 \\ 
                                           \Bg                                     &       2.71 &       4.06 &       5.87 &       8.81 &       6.84 &      10.26 \\ 
                                   \HH~1-0S(0)                                     &       0.38 &       0.57 &       0.43 &       0.64 &       0.52 &       0.78 \\ 
                                   \HH~2-1S(1)                                     &       0.27 &       0.40 &       0.31 &       0.47 &       0.31 &       0.46 \\ 
                                   \HH~1-0Q(1)                                     &       0.40 &       0.60 &       0.74 &       1.10 &       1.27 &       1.91 \\ 
                                   \HH~1-0Q(2)                                     &       0.33 &       0.50 &       0.44 &       0.66 &       0.52 &       0.78 \\ 
                                   \HH~1-0Q(3)                                     &       0.37 &       0.55 &       0.52 &       0.78 &       0.58 &       0.87 \\ 
                                          \Pfd                                     &      38.70 &      58.06 &      35.76 &      53.64 &      59.23 &      88.85 \\ 
                                           \Ba                                     &      16.66 &      24.98 &      19.82 &      29.74 &      31.50 &      47.24 \\

 				       [ArIII]                                     &\multicolumn{2}{c}{$\le$12}  &\multicolumn{2}{c}{$\le$11}  & \multicolumn{2}{c}{$\le$12} \\

                                         [SIV]                                     &\multicolumn{2}{c}{$\le$6}   &\multicolumn{2}{c}{$\le$5}   & \multicolumn{2}{c}{$\le$5}  \\ 

                                        [NeII]                                     &      110   &     170    &        440 &      660   &       570  &        850 \\ 
\hline
\multicolumn{7}{c}{$^a$spectral indexes $F\nu\propto\nu^{-\alpha}$}\\
                                                           $\rm\alpha^{6cm}_{2cm}$ &      -0.49 &      -0.29 &      -0.87 &      -0.67 &      -0.54 &      -0.34 \\ 
                                                          $\rm\alpha^{20cm}_{6cm}$ &      -0.29 &      -0.09 &      -0.66 &      -0.46 &      -0.03 &       0.17 \\ 
\hline

\end{tabular}
\end{center}
$^a$ spectral indexes from \citet{Sandqvist95}. An uncertainty of $\pm$0.1 is added.
\end{table*}

\subsection{Analysis of the images and registration} 
\label{Registration}

Fig.~\ref{fig colour}~displays a three colour image (0.6\micro,
2.2\micro~and 3.8\micro) of the central region of NGC1365. The
northern dust lane appears in green (2.2\micro), while the three
sources of interest show up in red.  Fig.~\ref{fig summary} provides a
summary overview of the imaging data. The large image presents the
central 30\arcsec$\times$30\arcsec~of NGC1365 through the HST F606W
filter (referred to as an R image since it is close to the standard R
band filter). This image is the pipeline-reduced HST/WFPC2 archive
image. The location of the bright type 2 AGN \citep{Lindblad99} is at
relative coordinates (0\arcsec; 0\arcsec). The square to the North of
the AGN outlines the region where the three bright MIR/radio~sources,
M4, M5 and M6, are found. The series of small images offers closer
views of this region at wavelengths: 0.6\micro~(WFPC2),
2.2\micro~(ISAAC), 3.8\micro~(ISAAC) and 12.8\micro~(VISIR), as well
as a reprint of the 3\,cm ATCA image from \citet{Forbes98}. On these
closer views, three pairs of concentric circles pinpoint the positions
of M4, M5 and M6, and the aperture sizes used for the flux density
measurements presented in Sect.~\ref{Measurements}.

The morphology of the nuclear region of NGC1365 varies with
wavelength. Hence, the relative registration of images at different
wavelengths has to be performed with care. In the following, we do not
provide absolute source positions but rather derive relative
registrations. For image orientations and pixel sizes, we use the
values attached to the image headers.  On the R, J and Ks band images,
enough sources are detected in the field of view to allow good
relative positioning. For this, we use the sources labelled C1, C2 and
C3 in Fig.~\ref{fig summary} which lie outside zones of highest
extinction. We do not use the AGN itself to register these images, as
a shift between its visible and infrared peak-positions may occur. On
the other hand, on the L' and N band images, only the AGN and the
sources M4, M5 and M6 have a S/N ratio sufficient to perform
positional measurements. In this case, the AGN must be used for
registration and we make it coincide with its location on the Ks band
image (Fig.~\ref{fig registration}). On the grounds of modelling, this
is an acceptable assumption as only a minor offset, if any, is to be
expected for the AGN position between the Ks and N bands
\citep{Granato97}.

Under such a registration, Fig.~\ref{fig registration} ascertains that
the positions of the reference sources C1, C2 and C3 agree within
better than $\pm$0.1\arcsec~on the R, J and Ks band images. At longer
wavelengths, as discussed above, one must rely on the location of the
AGN and, assuming that offsets of the nuclear peak projected positions
in the Ks, L' and 12.8\micro~ bands are small compared to the image
resolutions, the global precision on the registration of all maps is
better than $\pm$0.1\arcsec.  Estimate of the precision in the
relative positioning does matter since, as shown on the three bottom
rows of Fig.~\ref{fig registration}, the positions of the peaks
corresponding to M4, M5 and M6 appear to move slightly with
wavelength. In the following, we call ``location'' of M4, M5 or M6 the
projected position of their corresponding 12.8\micro~peak, which are
provided in table~\ref{table measurements}. They correspond to
projected distances to the AGN of 640\,pc, 920\,pc and 760\,pc for M4,
M5 and M6 respectively.

In the R and Ks bands, no conspicuous emission is detected at the
exact location of M4. Notice that M4 is located close to, but not
coincident with, the apex of a cone-shaped structure appearing on the
R band image, roughly extended along the North-East direction. This
structure is also bright in the J band. Emission in the Ks band is
detected between the location of M4 and the peak in the J band
mentioned above. It may be a mixture of emission from M4 and from the
``cone''. The source M4 is clearly detected in the L' band as well as
at 12.8\micro.

The source M5 is detected in the R, J, Ks and L' bands and at
12.8\micro.

The source M6 is detected in the Ks, L' bands, as well as at
12.8\micro.  Possible counterparts to M6 are detected, slightly offset
to the South, in the R band (offset by 0.2\arcsec) and J band (offset
by 0.1\arcsec).

Regarding the 3\,cm~image, we only have access to the printed figures
by \citet{Forbes98} and \citet{Morganti99}, therefore no precise
measurement can be performed. Fig.~\ref{fig MIR-cm} presents the
filtered 12.8\micro~image with cuts highlighting the extended emission
and compares this image to the 3.5\,cm map by \citet{Morganti99}. The
similarity of the two maps is striking, not only for the main peaked
sources, but also for the extended emission and in particular for the
western feature elongated in the North-East direction.  The sources
M4, M5 and M6 appear to be distributed in a well defined star-forming
ring, delineated through its MIR/radio emission. The AGN itself is
neither radio bright, nor clearly isolated on the radio image, and
hence cannot be used for registration purposes. The two independent
maps by \citet{Forbes98} and \citet{Morganti99} are in good
agreement. Both indicate that the three radio sources cannot be made
precisely coincident with the three MIR sources simultaneously. In
Fig.~\ref{fig summary}, M6 is chosen as the reference: M4 and M5 are
slightly shifted to the North-East with respect to the radio map. In
Fig.~\ref{fig MIR-cm}, we display an alternative solution in which M4
and M5 have precise radio counterparts: then, M6 appears to be
slightly shifted to the South-West with respect to the radio
emission. As each of the two independent radio maps shows this offset
--in its comparison with the MIR image--, one is led to conclude that
the offset is real, but remains to be understood.

The sources M4 and M6 appear to be located on the dust lane, in
projection. The fact that, in the R band, M4 is not detected
  and M6 is very weak suggests that they lie inside the dust
  lane. The source M5 lies on the edge of the dust lane, and
consistently appears less extincted, showing a bright counterpart in
the R band. Halos of emission are detected around M4, M5 and M6, both
in the L' band and at 12.8\micro.

In conclusion, the relative positioning of the visible and IR images
has been achieved at a precision of 0.1\arcsec and is adopted in the
following analysis.

Let us make some cautionary remarks in the case of M4: the comparisons
between its maps indicate that aperture flux measurements should be
unreliable in the R, J and Ks bands, where no clear counterparts are
found and as it lies in the vicinity of an R band emitting cone and of
a J to Ks band emitting region. We believe that, at these wavelengths
(R, J and Ks bands), aperture flux measurements for M4 are upper
limits.

The source M5 is well defined at all wavelengths. A neighbouring weak
source to the North (seen on the R band image) may contribute some
flux to its aperture flux measurements in the J and Ks bands. Yet, M5
probably remains the dominant source in the Ks band.

The source M6 is clearly detected at all wavelengths except in the R
band. At this wavelength, a weak source is detected
$\sim$0.2\arcsec~to the South of the location of M6 (See Fig.~\ref{fig
  summary} and Fig.~\ref{fig registration}). We cannot firmly exclude
that it is an unrelated source, therefore its flux has been included
in computing the R band flux upper limit of M6.

These remarks of course apply to continuum measurements from the
spectra that we are about to discuss, since they correspond to a
1\arcsec$\times$1.4\arcsec~aperture. This leads to the following
conclusion, to be kept in mind when interpreting the NIR spectra: in
the case of M4, the Ks band continuum is an upper limit, in the case
of M5 it may be slightly overestimated, and in the case of M6 it is a
proper estimate. At longer wavelengths, such issues are irrelevant.

\subsection{Analysis of the spectra}
\label{Spectra} 

In the upper part of Fig.~\ref{fig spectra}, we provide drawings of
the slits used for the spectroscopy in the Ks and L' bands,
superimposed on images in the R and Ks bands and on the
12.8\micro~image. Along the slits, thin ticks are drawn every arcsec,
while longer and thicker ticks highlight the positions at which
emission line spectra have been recorded. References used for the
positioning of the spectra along the slits are the AGN for slit 1 and
M6 for slit 2. M6 is a good reference since its position is well
defined and does not shift significantly with wavelength above
1\micro.

For M4, the emission line spectrum position (given by the thick tick
mark on the slit drawing) coincides with the emission peak on the
12.8\micro~image; it is located 0.1-0.2\arcsec~to the South of an
emission knot in the Ks band (but has itself no clear counterpart in
Ks); it is also located 0.2\,\arcsec~to the East of the apex of the
cone-shaped structure seen in the R band image.

The emission line spectrum at the position of M5 corresponds to well
defined emission knots on the three images (R, Ks and 12.8\micro).

The emission line spectrum at the position of M6 corresponds to an
emission knot both on the 12.8\micro~and Ks images, but is
0.1-0.2\,\arcsec~ to the North of an emission knot in the R image.

The 2D raw spectra obtained through slit 2 in the Ks and L' bands with
ISAAC, and in the N band with VISIR, are displayed at the bottom of
Fig.~\ref{fig spectra}. The brightest emission lines identified on the
figure are \Pa~and \Bg~in the Ks band, \Ba~in the L' band and [NeII]
in the N band.

The extracted spectra are displayed in Figs.~\ref{fig spectrum
  M4},~\ref{fig spectrum M5}~and~\ref{fig spectrum M6} for M4, M5 and
M6 respectively. The following set of emission lines is detected: (a)
nebular lines \Pa, \Bd, \HeI, \Bg, \Pfd, \Pfg~, \Ba, [NeII]
12.8\micro, (b) molecular lines from \HH~and (c) PAH emission. We
detect PAH signatures at 3.3\micro~and 11.3\micro. The bump-like
feature on the 8-9\micro~spectrum of M6 might be a signature of the
8.6\micro~PAH, although it looks a bit broad. Even though the
[NeII]12.8\micro~line is bright, we do not detect any
[ArIII]8.9\micro~or [SIV]10.5\micro~line emission, which is a puzzling
and interesting result.  The 12.7\micro~PAH feature, expected to show
up as a broad feature, is also absent. For the three sources, the
continuum spectral distribution is flat or slightly rising with
wavelength in the Ks band and neatly rising in the L' and N bands. The
presence of the 9.7\micro~silicate absorption band is inferred in M4,
M5 and M6, from the simultaneous fall and rise of the
8.1-9.3\micro~and 10.4-13.0\micro~continua respectively matching the
blue and red wings of the silicate absorption feature. For the
continuum in the Ks band, one must bear in mind the remarks given at
the end of the previous section.  Between 3\micro~and 3.3\micro~, the
transmission of the atmosphere is poor: this part of the spectrum is
noisier and we think that the uncertainty on the atmospheric
correction induces a supplementary error on the continuum value.  We
are tempted to believe that only the segment of the spectrum red-wards
of 3.3\micro~is reliable. In the N band, the spectral calibration for
the red side of the spectrum has been performed through scaling to the
VISIR narrow band 12.8\micro~image and taking advantage of the
overlapping region between 11\micro~and 12\micro. This procedure
resulted in applying scaling factors to the spectra. Such a correction
cannot be applied to the blue side of the N band spectrum, since no
overlapping spectrum is available. Therefore, the relative scaling
between the 8.1-9.3\micro~segment and the 10.4-13\micro~segment of the
spectrum may be uncertain by a factor up to 2.

On the spectra, an interesting feature is the extension of the line
emission along the slit direction (see the Ks band spectrum in
Fig.~\ref{fig spectra}) for the brightest lines. For M4, and in spite
of a rather low spectral resolution, we can even detect a velocity
gradient in the \Pa~line. The position-velocity diagram for the M4
\Pa~emission line is displayed in Fig.~\ref{fig PV}. Between positions
-1\arcsec~and +1\arcsec, a velocity difference of the order of
100-200\,\kms~is measured. This kinematical feature will be discussed
elsewhere. In the case of M5 and M6, we also detect extended nebular
emission (see Fig.~\ref{fig spectra}), but no velocity gradient.

Spectral analysis in the radio domain would also be of great
interest. The radio measurements by \citet{Sandqvist95} indicate that
the spectral indexes in the three sources flatten with increasing
wavelength, suggesting optically thick radio emission. Unfortunately,
such measurements are delicate from an image with complex brightness
distribution and have not been repeated by \citet{Morganti99}, hence
cannot be double checked.

In summary, the collected spectra exhibit bright nebular and molecular
emission lines at the location of the MIR/radio sources M4, M5 and M6:
this reinforces the idea that these sources are embedded YMCs. The
line emission is spatially extended; in the case of M4, a velocity
gradient over a few arcsec is detected, suggesting the presence of an
outflow.

\subsection{Measurements and related uncertainties} 
\label{Measurements}

Table~\ref{table measurements} displays the full set of measurements
performed for the three sources M4, M5 and M6.  The flux density
measurements were made through apertures of 0.6\arcsec~radius,
represented in Fig.~\ref{fig summary} by the small inner circles, and
centred on the source positions as determined on the VISIR
12.8\micro~image. Fluxes were measured on the following images: the
HST/WFPC2 R band image, the ISAAC J, Ks, L' band images, the TIMMI2
10.4\micro, 11.9\micro~and 12.9\micro~images \citep{Galliano05a}, and
the VISIR 12.8\micro~image. In \citet{Galliano05a}, we suspected the
12.9\micro~flux densities to be overestimated by a factor 2. The
TIMMI2 N band spectrum allows a direct measurement of the
11.9\micro/12.9\micro~flux ratio (independent of the flux
calibration), and shows that our suspicion was founded
indeed. Therefore, the 12.9\micro~fluxes for NGC1365 quoted in
\citet{Galliano05a} must be decreased by a factor 2.

Around each aperture, a ``background'' level was estimated by
computing the median pixel value in an annulus with radii
0.6\arcsec~and 1.2\arcsec~around the aperture centre. The large and
outer circles on Fig.~\ref{fig summary} represent the outer borders of
such annuli. We cannot be certain that this ``background'' also
affects the source within the small aperture, as its origin and its
location with respect to the source are unknown. Therefore, in a
conservative approach we provide two flux density measurements for the
sources M4, M5 and M6: one after ``background'' subtraction and one
without ``background'' subtraction. The two figures for each flux
represent the measurement uncertainty resulting from an ambiguity in
the interpretation. To this uncertainty we add quadratically the
uncertainty due to the photometric calibration.  The calibration
uncertainties used for the different bands are as follows, R: 10\%, J:
10\%, Ks: 20\%, L': 20\%, 10.4\micro: 20\%, 11.9\micro: 20\%,
12.9\micro: 20\% and 12.8\micro: 20\%. Whenever at a given wavelength
no counterpart is clearly identified at the position of the MIR
source, then we consider the high value of the error bar as an upper
limit. In table~\ref{table measurements}, the low and high
  values given for each measurement correspond to the lower and upper
  limits for the given measurement due to the uncertainties defined
  above.

Table~\ref{table measurements} also provides the flux measurements for
the emission lines identified in the spectra of M4, M5 and M6. In the
case of M6, we have additional measurements for lines in the N
band. Finally, we recall in this table the radio measurements
published by \citet{Sandqvist95}.  Fig.~\ref{fig SED} displays, for
the three embedded YMCs, their SED together with their combined
spectrum, covering in total a range from 0.5\micro~to 13\micro. Let us
briefly discuss the comparison between SEDs and spectra. The SED
points show flux density values that we believe correspond directly to
the sources of interest. However, in some cases they may correspond to
upper limits because of source confusion. On the contrary, the spectra
contain effective signal from the source and its surrounding. This
distinction is particularly relevant in the case of M4. That is why on
Fig.~\ref{fig SED}, for M4 the Ks band spectrum is shown as an upper
limit. For the sources M5 and M6, the Ks band continuum corresponds to
the upper part of the error bar on the flux measurement, because this
value corresponds to the measurement performed without ``background''
subtraction. The L' band data points and the N band data points of
\citet{Galliano05a} are in excellent agreement with the spectra. The
TIMMI2 10.4\micro~flux density for M4 is not reported on this figure,
since there was no detection on the TIMMI2 image and the, then quoted,
upper limit of 5mJy may have been underestimated.

\section{Comparison with a resolved young massive cluster: R136 in the LMC}
\label{R136}
\begin{figure}[htbp]
\begin{center}
\includegraphics[width=9cm]{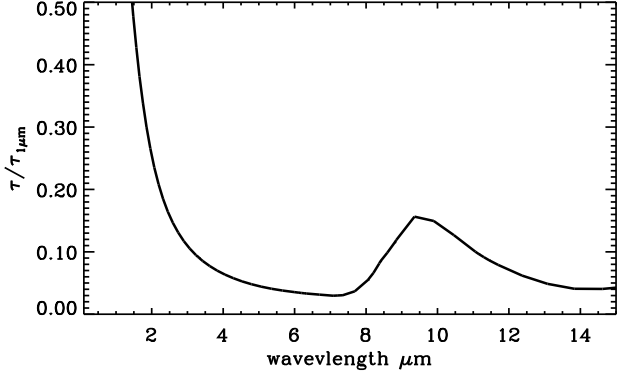}
\caption{Extinction curve used in this paper. This curve has been derived with \texttt{GRASIL} for the average Galactic extinction \citep{Draine84}.}
\label{fig extinction curve}
\end{center}
\end{figure} 
\begin{figure*}[htbp]
\begin{center}
\includegraphics[width=12cm]{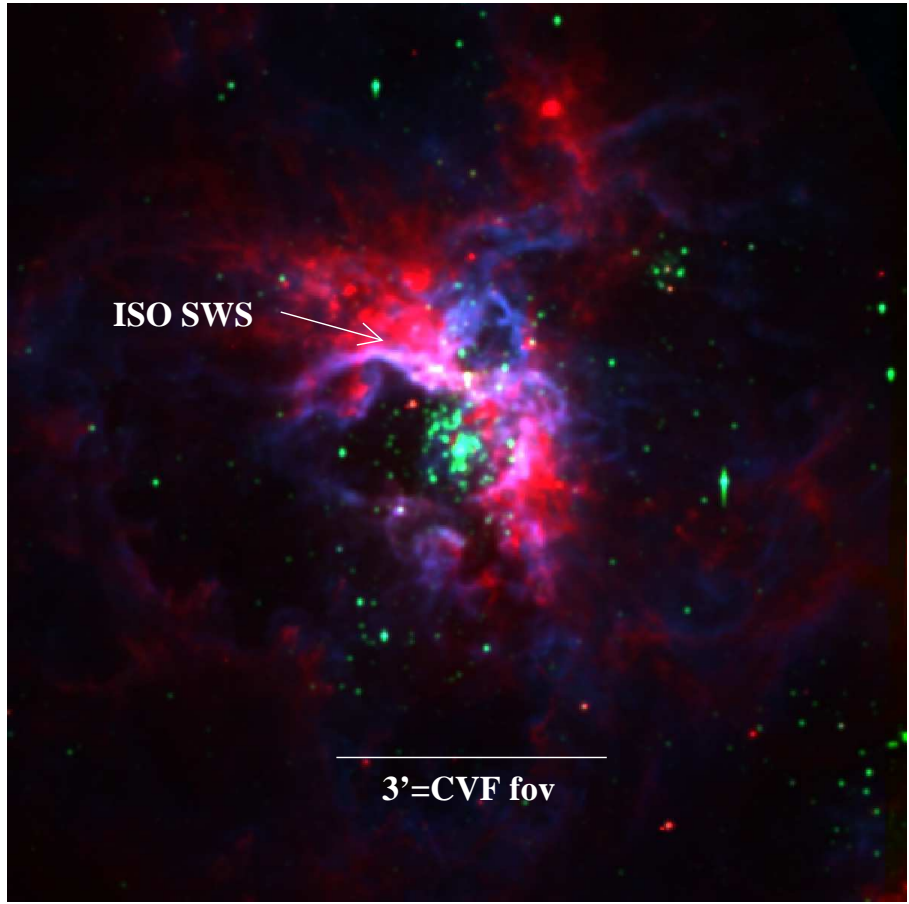}
\caption{ESO 2.2m WFI/Spitzer IRAC view of R136 in the Large Magellanic Cloud. The field of view is 10\arcmin$\times$10\arcmin. North is up and East is to the left. This region (145pc$\times$145pc) would project onto a field of 1.6\arcsec$\times$1.6\arcsec~at the distance of NGC1365. Blue codes the WFI \Ha~image, green the WFI V image and red the IRAC 8.0\micro~Spitzer image. The 3\arcmin~scale corresponds to the aperture size of the CVF spectrum aperture and the arrow shows the location of the 14\arcsec$\times$20\arcsec~ISO SWS spectrum aperture.}
\label{R136 image}
\end{center}
\end{figure*} 
\begin{figure}[htbp]
\begin{center}
\includegraphics[width=9cm]{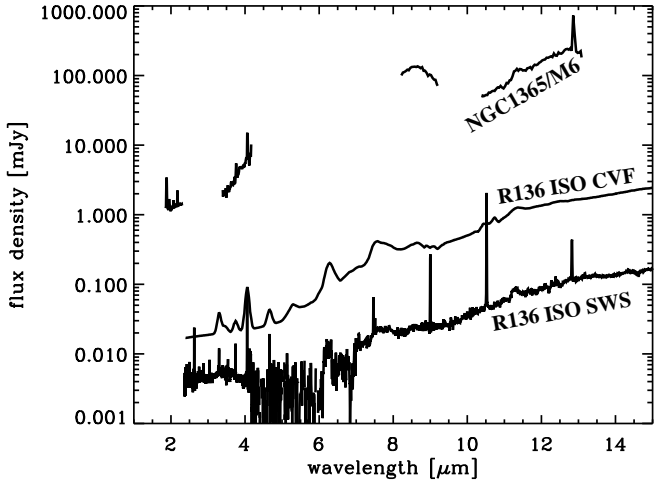}
\caption{Comparison of the ISAAC/VISIR spectrum of NGC1365/M6, the ISO CVF spectrum of R136 and ISO SWS spectrum of R136. The spectra of R136 are projected to the distance of NGC1365.}
\label{comp spec R136 M6}
\end{center}
\end{figure} 
\begin{figure*}[htbp]
\begin{center}
\includegraphics[width=12cm]{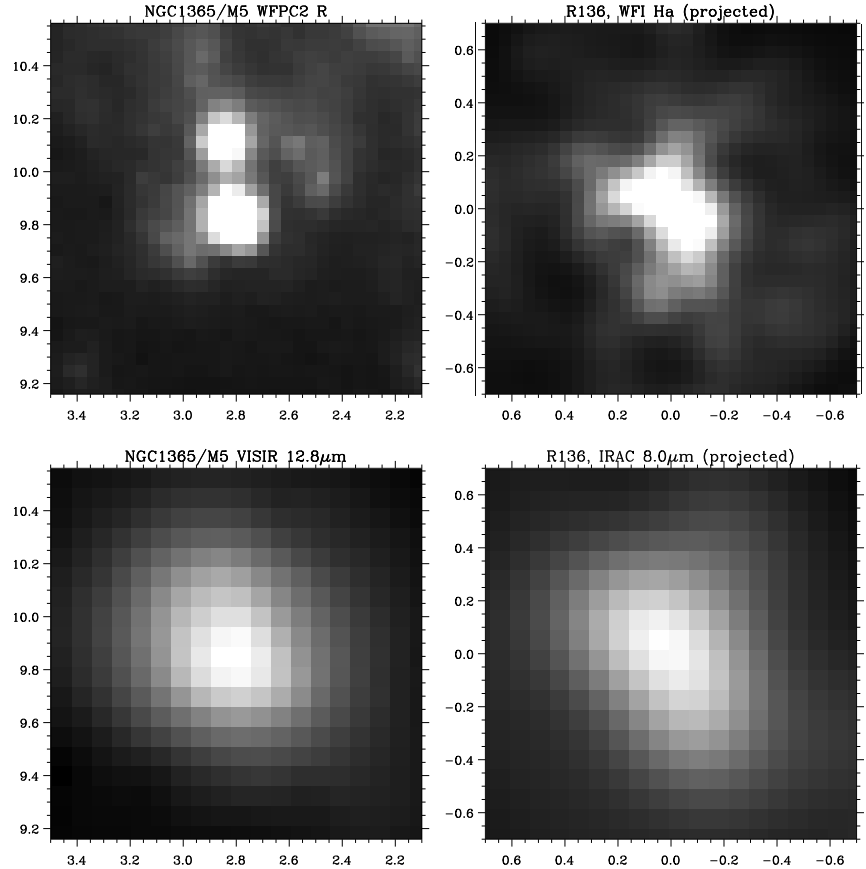}
\caption{Map comparison between NGC1365/M5 (left column) and R136 projected at the distance of NGC1365. The R136 WFI \Ha~image and IRAC 8\micro~image, after projection at the distance of NGC1365, have been degraded to the seeing and pixel size of the NGC1365/M5 WFPC2 R image and the VISIR 12.8\micro~image respectively.}
\label{R136 ima comp}
\end{center}
\end{figure*} 
\begin{figure}[htbp]
\begin{center}
\includegraphics[width=9cm]{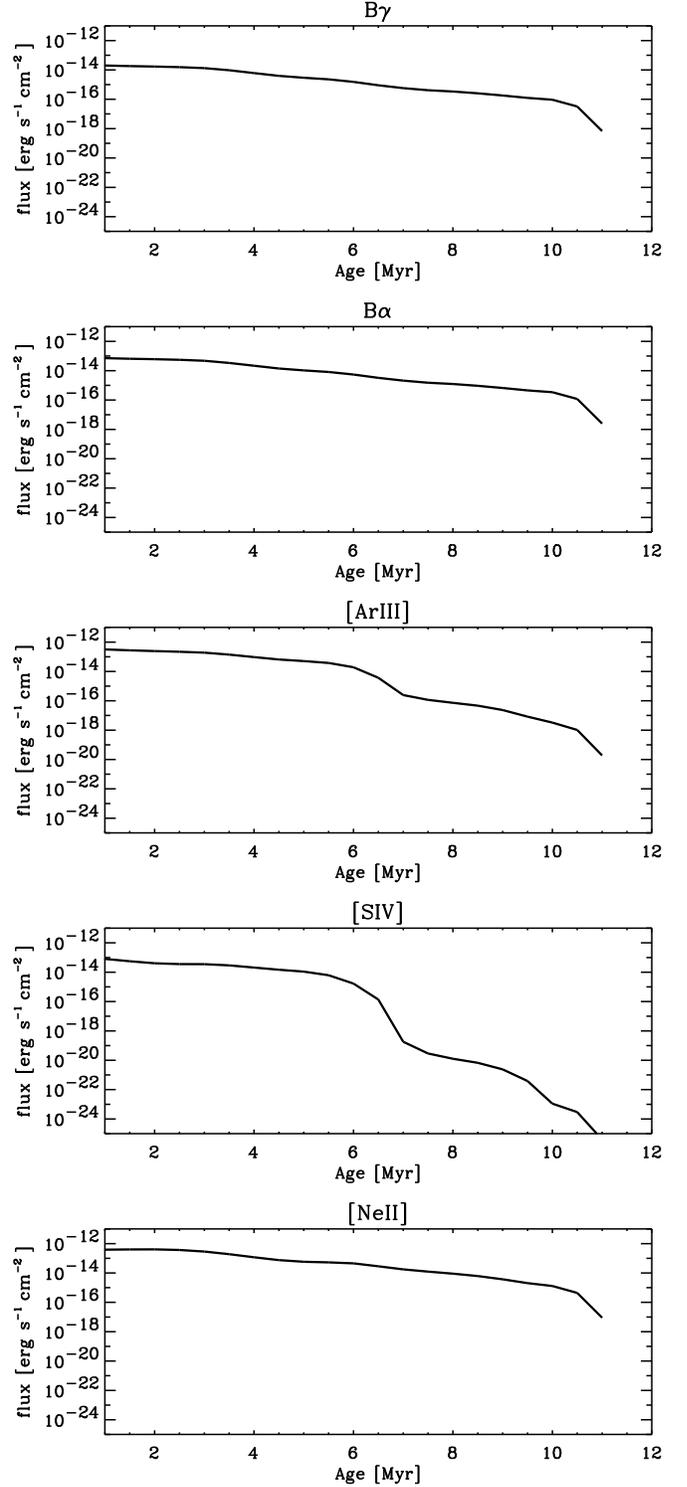}
\caption{Predicted line fluxes for the model described in Sec.~\ref{Age and mass}. From top to bottom: \Bg, \Ba, [ArIII], [SIV] and [NeII]. The X-axis displays the age in Myr and the Y-axis provides the line flux, at the distance of NGC1365, corresponding to a 10$^6$\msol~cluster}
\label{fig line fluxes}
\end{center}
\end{figure} 
\begin{figure*}[tbp]
\begin{center}
\includegraphics[width=12cm]{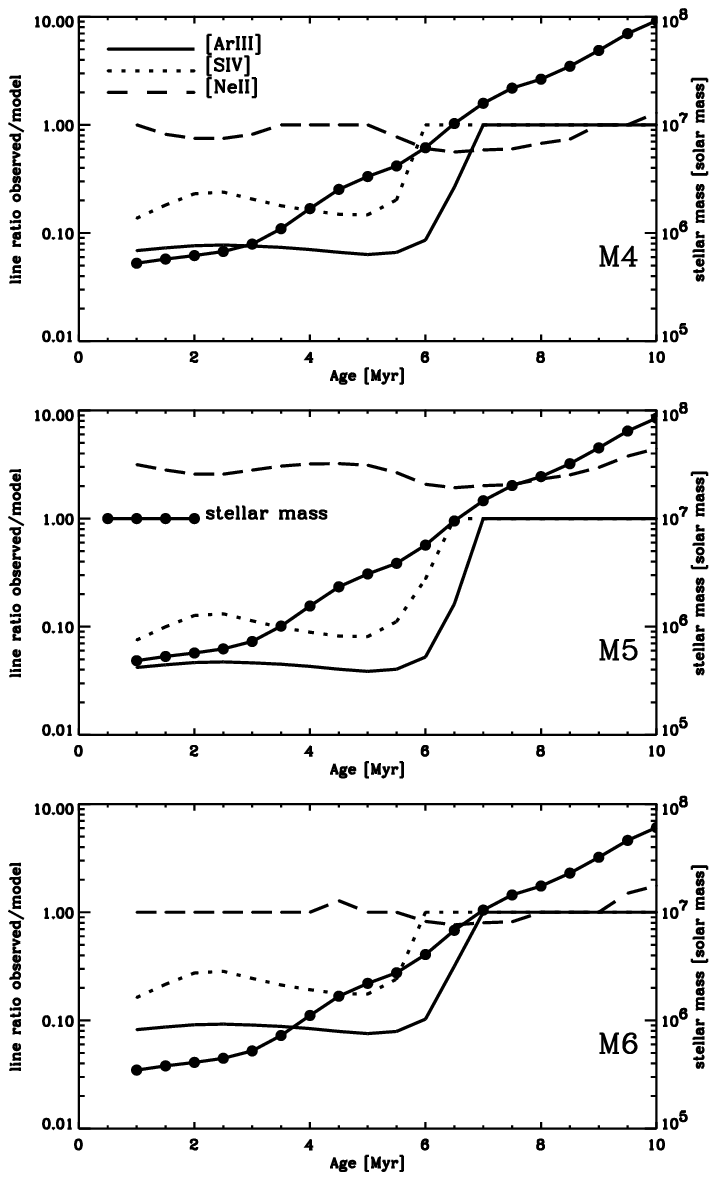}
\caption{For each cluster, and for the extinction derived in Sec.~\ref{extinction}, this figure gives the ratios between observed and modelled [ArIII], [SIV] and [NeII] lines after correcting for the extinction correction and normalising the modelled \Bg~fluxes to the observed ones. The ratio is set to one when the modelled flux is inside the error bar, or below the upper limit, which is the case for [ArIII] and [SIV]. The total stellar mass (taking into account evolutionary effects) is also plotted for each possible age.}
\label{line ratios}
\end{center}
\end{figure*} 


The fact that the MIR/radio sources M4, M5 and M6 in NGC1365 exhibit
nebular lines strongly supports an interpretation in terms of embedded
YMCs still hosting a fair fraction of ionising stars, hence younger
than 10\,Myr.

Let us start the analysis by comparing the available data for the YMCs
in NGC1365 with similar information (images and spectra) on R136, a
young cluster in the LMC. Basic parameters for these clusters will be
derived in subsequent sections.
 
R136 is one of the nearest example of a young massive cluster. It lies
in the most luminous star forming region in the Local Group: 30
Doradus in the LMC, at a distance of 50\,kpc. R136 has an age in the
range 1-3\,Myr and a stellar mass of
5$\times10^4$\msol~\citep{Boulanger06}.

Even though R136 is not properly speaking an embedded cluster, it is
an interesting object for comparing with the MIR/radio emitting
sources of NGC1365. A three colour map of R136 is displayed in
Fig.~\ref{R136 image}, built as a composite of two archived images
obtained with the Wide Field Imager (WFI) at the 2.2m telescope on La
Silla (ESO) and one archived image collected with IRAC on board the
Spitzer space telescope (NASA). Green codes the WFI V-band image
(tracer of the stellar component), blue codes the WFI \Ha~image
(tracer of the ionised gas) and red codes the IRAC 8.0\micro~image
(tracer of dust). The resolution of the WFI images has been degraded
to 3.2\arcsec, which is the resolution of the IRAC
8.0\micro~image. The composite field size is
10\arcmin$\times$10\arcmin, which corresponds at the distance of the
LMC to 145$\times$145pc$^2$. This field projects into a region of
1.6\arcsec$\times$1.6\arcsec~at the distance of NGC1365.  This image
shows that, even if R136 does not qualify as ``embedded'' cluster, it
is a bright source of MIR radiation: would it lie behind or within an
extinction lane, it would indeed be identified as an ``embedded''
cluster. This suggests that the MIR emitting dust of a so-called
``embedded'' cluster may not be necessarily in direct relation with
the ``embeddedness'' of the cluster, as assumed in the typical
cocoon-like view of an embedded cluster. As highlighted on this colour
image, the three components are spatially separated: one can clearly
identify the \Ha~emitting gas, spatially correlated with but distinct
from the MIR dust component, and the stellar cluster itself. The wide
dust shell (bubble) pushed by the cluster is prominent, while the
\Ha~emission shows up as a thin layer at its inner rim.

Of the many spectra available for R136, let us consider: (i) the ISO
SWS spectrum of a 14\arcsec$\times$20\arcsec~region in the North-East
MIR emitting lobe (shown by an arrow on Fig.~\ref{R136 image}) and
(ii) the ISO CVF spectrum of the 3\arcmin$\times$3\arcmin~region
around the central cluster, covering the wavelength range from 2 to
16\micro~\citep[taken from Fig.~3 of ][]{Boulanger06}.

Notice that an aperture of 3\arcmin$\times$3\arcmin~in the LMC
translates to 0.5\arcsec$\times$0.5\arcsec~at the distance of
NGC1365. This allows a direct comparison between the CVF spectrum of
R136 and the ISAAC/VISIR spectra of the embedded YMCs in
NGC1365. Fig.~\ref{comp spec R136 M6}~displays the SWS and CVF spectra
of R136 projected at the distance of NGC1365, as well as the spectrum
of N1365/M6. This comparison shows that M6 is, intrinsically, almost
two orders of magnitude brighter than R136 and that, given the
available data, the two sources exhibit at first order a rather
similar spectrum. A remarkable difference, though, is the fact that in
the 8-13\micro~wavelength range, three bright emission lines are
detected on the SWS spectrum of R136 ([ArIII] at 8.9\micro, [SIV] at
10.5\micro~and [NeII] at 12.8\micro) while only the [NeII]
12.8\micro~line is observed in the case of M6. This difference holds
true in the case of M4 and M5. Owing to its low spectral resolution
the R136 CVF spectrum does not show any of the narrow forbidden lines
but only broad features such as the PAH bands.

We provide in Fig.~\ref{R136 ima comp}, a comparison of the WFPC2 R
and the [NeII]12.8\micro~images of M5 with the WFI \Ha~and the IRAC
8.0\micro~images of R136 projected at the distance of NGC1365, with
same seeing and same pixel size as the M5 images. For this comparison
we have chosen M5 since, among the three YMCs in NGC1365, it is the
one with the brightest visible counterpart, hence the closest to the
evolutionary stage of R136. On the left side of Fig.~\ref{R136 ima
  comp}, the images of M5 are displayed, while on the right side, the
corresponding degraded images of R136 are shown. Notice that the bulk
of the 8.0\micro~emission in R136 looks elongated, while its
counterpart in M5 is unresolved. The same occurs with the visible
image: at the resolution of HST, the degraded R136 looks slightly more
extended than M5.  This comparison suggests that, in terms of
intrinsic properties, the YMCs in NGC1365 are slightly smaller than
R136 while they are about two orders of magnitude brighter: they
definitely deserve to be called ``compact''. Another piece of evidence
comes from the HST image, where M5 remains unresolved, at a resolution
of 0.11\arcsec, which corresponds to about 10\,pc.  The fact that the
sizes of the YMCs in NGC1365 are comparable to, or even less than,
that of R136, which is 100 times fainter, supports the idea that star
cluster sizes do not depend on mass \citep{Larsen04}.

This simple and direct data comparison between the MIR/radio sources
in NGC1365 and R136 adds support to the interpretation of the NGC1365
sources in terms of compact YMCs.

An additional striking difference between the two sources is the
absence of the [ArIII] and [SIV] line emission in the YMCs in
NGC1365. This important difference is discussed in Sec.~\ref{YMC basic
  parameters} and explained through an age effect, the clusters in
NGC1365 being older than R136.  A detailed comparison of the SEDs of
R136 and of the YMCs in NGC1365 is performed and included in
Sec.~\ref{SED}.

\section{Derivation of the YMC basic parameters}
\label{YMC basic parameters}
In this section, we attempt to derive parameters for the YMCs in
NGC1365, such as extinction, age and mass, using their emission line
flux measurements. We make the gross assumption that a uniform
foreground emission is affecting in a similar way all the emitting
components in the cluster. This is obviously an oversimplification,
but it leads to first-order interesting conclusions while avoiding to
use model-fitting with too many free parameters.  We shall turn to a
more realistic modelling in the next
section.

\subsection{Extinction}
\label{extinction}
First, let us use the \Bg~and \Ba~line measurements to derive the
extinction towards the YMCs. We know already that they are located in
a region of large extinction, in the vicinity of, or within the
prominent East-West dust lane to the North of the AGN (Fig.~\ref{fig
  summary}). In the following discussion, we use the extinction curve
derived with \texttt{GRASIL} \citep{Silva98} for Galactic dust. This
curve is shown in the NIR/MIR range in Fig.~\ref{fig extinction
  curve}.

A direct estimate of the extinction towards the embedded cluster
nebular gas can be retrieved from the \Bg/\Ba~line ratio, assuming
optically thin conditions for the nebular gas itself. This ratio
depends only mildly on the gas density.  Moreover, considering that
the temperature of HII regions tends to cluster around $\rm
T_e=10^4\,K$ \citep{Osterbrock89}, the predicted line ratio of
\Pa~(1.87\micro) to \Bd~(1.94\micro) is 0.0545.  The differential
reddening between \Pa~and \Bd~is small, the two lines being quite
close in wavelength, so we can use the directly measured \Pa/\Bd~ratio
to check the conditions in the gas, and the validity of assuming the
case-B line ratios. For the embedded clusters M4, M5 and M6, the
measured ratios are respectively 0.053, 0.053 and 0.058, in good
agreement with the theoretical value 0.0545 mentioned above.

Consequently, the case-B assumption looks adequate and we can
confidently derive the extinction towards the embedded cluster nebular
gas using the theoretical \Bg/\Ba~flux ratio of 0.35
\citep{Osterbrock89}.  For M4, M5 and M6, we obtain NIR extinctions of
A$\rm _V$=13.5, 3.2 and 8.5, respectively.We use these values to
de-redden the NIR line fluxes.

\subsection{Age and mass}
\label{Age and mass}
Together with \Bg~and \Ba, we can use the measurements of the MIR fine
structure lines [ArIII]\,8.9\micro, [SIV]\,10.4\micro~and
[NeII]\,12.8\micro~in order to derive the cluster
parameters. Predicting the intensities of the fine structure lines is
more complex than in the case of the hydrogen lines. Their ionisation
potentials are larger (respectively 27.6eV, 34.8eV and 21.6eV), which
implies that not only the H ionising luminosity, QH, must be taken
into account, but also that the shape of the ionising continuum must
be known or assumed. Moreover, the fluxes in these lines depend on the
ionisation factor, closely linked to the geometry of the source while
we do not have much handle on it. Finally, the line fluxes depend on
the element abundance ratios: given the location of the YMCs in their
host galaxy, a set of solar abundances looks adequate even though a
slight over abundance (up to a factor two) could be expected.

What is remarkable on the spectra of the three sources, M4, M5 and M6,
is that while the [NeII] line is conspicuous, only upper limits can be
measured for the [ArIII] and [SIV] lines: the [NeII]/[SIV] line ratios
are greater than 20, 100 and 140 for M4, M5 and M6 respectively, and
the [NeII]/[ArIII] line ratios are larger than 10, 50 and 60. Such
ratios are impossible to reproduce with very young stellar
populations, for which the line ratio [NeII]/[SIV] is usually found to
be lower than unity \citep[see the SWS spectrum of R136 and for
  instance the spectrum of NGC5253 in ][]{Martin-Hernandez05}. So, a
substantial difference in the line ratios is unveiled, by one to two
orders of magnitudes.

In order to examine this difference, we use the \texttt{CLOUDY}
emission line libraries computed by
\citet{Panuzzo03}\footnote{http://web.pd.astro.it/panuzzo/hii/index.html}
for branching into \texttt{GRASIL} \citep{Silva98}. In these
libraries, the ionisation factor is varied by changing the gas density
and the gas filling factor. \citet{Panuzzo03} demonstrate that
different geometries with a similar final ionisation factor will
produce very similar line fluxes.  We compute the \Bg, \Ba, [ArIII],
[SIV] and [NeII] line fluxes for a ``cluster model'' at ages between 1
and 10\,Myr with an ''instantaneous'' star formation history.  As
mentioned above, we adopt solar abundances and we notice that in no
way an overabundance by a factor two could account for the very large
observed differences in line ratios.

Since the [ArIII]/[NeII] line ratios observed in M4, M5 and M6 are
very small, we use the libraries predicting the smallest values for
this ratio: filling factor of $10^{-3}$ and density of
10$^4$cm$^{-3}$. The evolution with age of the un-extincted line
fluxes, at the distance of NGC1365, are displayed in Fig.~\ref{fig
  line fluxes}, for a total mass of stars of 10$^6$\msol. At first
order, these line fluxes show, for reasonable values, little
dependence on the amount of gas, in agreement with the
ionisation-bound situation.

In Fig.~\ref{line ratios}, we present a comparison between the
modelled line fluxes and the observed ones. The procedure for building
this figure for the three YMCs in NGC1365 was the following:
\begin{itemize}

\item the extincted modelled line fluxes are computed, for each YMC,
  with the related extinction value derived above,

\item the extincted modelled line fluxes are then normalised, by
  changing the cluster mass, so that the modelled \Bg~line flux equals
  the observed \Bg~flux, 

\item the line ratios between observed and modelled line fluxes are
  then computed for each age and plotted (left Y-axis) as a function
  of age. Whenever the modelled flux falls inside the observational
  error-bar or is lower than the detection upper limit, then the ratio
  is set to unity,

\item for each age, the corresponding stellar mass is also plotted
  (right Y-axis).

\end{itemize}

The model can be considered to ``fit'' the data if all three ratios
are equal to one. This is absolutely not possible before 6\,Myr, where
we would expect intense [ArIII] and [SIV] lines, one order of
magnitude brighter than allowed by the upper limits measured on the
spectra of the three YMCs. A reasonable ``fit'' can be obtained only
for ages greater than 6\,Myr. Notice that in the case of M5, there
remains a small discrepancy, by a factor two, for the [NeII] flux,
which, given the large differences involved and the first-order
matches performed, is left aside.

\begin{figure}[htbp]
\begin{center}
\includegraphics[width=9cm]{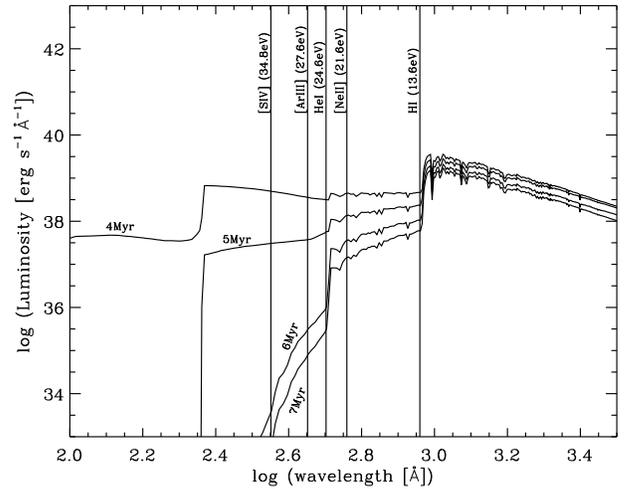}
\caption{Starburst99 model spectra for an instantaneously formed Salpeter cluster at ages 4, 5, 6 and 7\,Myr. The vertical lines show the first ionisation potentials for H, [NeII], He, [ArIII] and [SIV].}
\label{NeIIwithhands}
\end{center}
\end{figure}

In Fig.~\ref{NeIIwithhands}, we plot the Starburst99 spectra of an
instantaneously formed Salpeter star cluster at ages 4, 5, 6 and
7\,Myr, and compare them with the ionisation energy of [NeII], [ArIII]
and [SIV]. This figure shows , between 5 and 6\,Myr, a huge drop of
luminosity (of about two orders of magnitude) for the continuum in the
region of the ionisation energies for [ArIII] and [SIV]. On the
contrary, the continuum around the ionisation energy for [NeII] only
suffers a modest decrease. This shows that the virtual absence of
[ArIII] and [NeII] lines after 6\,Myr is due to the ``absence'' of
hard enough continuum for these ion species.

The firm conclusion to be drawn from Fig.~\ref{line ratios} is a
strong evidence for the YMCs in NGC1365 to be older than
6\,Myr. Because of the important fading suffered by stellar clusters
along the first Myrs of their evolution, the fact that they are
relatively old also implies that they are very massive. Let us
consider the age of 7\,Myrs: the computed stellar masses in this case
for M4, M5 and M6 are then respectively 1.6,1.5 and 1.0
$\times$10$^7$\,\msol~for a 1\msol~lower mass limit of the IMF. 

\begin{figure}[htbp]
\begin{center}
\includegraphics[width=9cm]{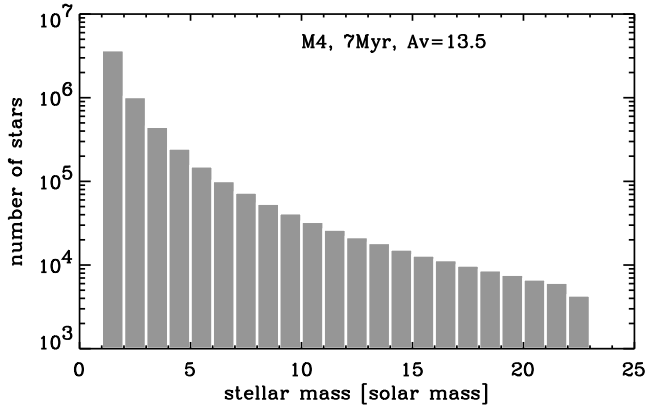}
\caption{Mass function for M4, given the age and mass derived in Sec.~\ref{Age and mass}.}
\label{M4 mass function}
\end{center}
\end{figure}

We show in Fig.~\ref{M4 mass function} the stellar mass function in M4
with these parameters (age of 7\,Myrs and mass of
1.6$\times$10$^7$\msol). The most massive stars in the clusters,
20-25\msol, are a few several $10^4$ in number, while low mass stars
are expected to be present by millions.

Consideration of a line flux library with lower filling factors shows
that, while the [ArIII] line flux is not affected, the [SIV] line flux
may increase by a factor up to 10. This remains compatible with the
data as long as the clusters are only about 1\,Myr older than the age
derived for a filling factor of $10^{-3}$. Hence, a safe range for the
age of the YMCs is 6-8\,Myr. We exclude the clusters to be older than
8\,Myr since if it was the case, the masses derived would be of the
order of 10$^8$\msol~for the only stellar component and become
incompatible with the CO measurements by \citet{Sakamoto07}.

This is consistent with the detection of the CO absorption lines at
2.3\micro~in the spectra of M4, M5 and M6, since this CO feature only
appears after 6\,Myr \citep{Leitherer99}.

In principle, one could also use the ratio HeI~2.06\micro/\Bg~to
derive information about the hardness of the ionising continuum
radiation, hence the age of the stellar population \citep{Doyon92}.
However, \citet{Shields93} and \citet{Lumsden03} have demonstrated
that the use of this ratio is not trustworthy.

\subsection{Ionising photon emission rate}
Using the NIR extinctions derived above, we can compute the
de-reddened hydrogen line fluxes, and get a direct handle on the
ionising photon emission rates of the YMCs.  For an ionisation-bound
situation, justified for an embedded cluster, all the ionising photons
($\lambda \le 912\AA$) emitted by the stars are absorbed. The number
of H ionising photons is then directly proportional to the flux in any
specific recombination line.  From \citet{Osterbrock89}:
\begin{equation}
\rm Q[H^+]=\frac{\alpha_B}{\alpha^{eff}_{H\alpha}} \times \frac{L_{H\alpha}}{h\nu_{H\alpha}}~~,~where~\alpha_B/\alpha^{eff}_{H\alpha}\sim 2.96
\end{equation}
The \Bg~de-reddened fluxes of M4, M5 and M6 being respectively in the
ranges [8.2-12.3], [7.6-11.5] and [13.7-20.6] $\times 10^{-15}$\flux,
this leads to ionising photon emission rates in the ranges [3.4-5.1],
[3.2-4.8] and [5.7-8.6] $\rm \times 10^{52} s^{-1}$.

Let us check now whether the ionising photon emission rates derived
from the \Bg~de-reddened fluxes of M4, M5 and M6 are consistent with
the observed radio data in the centimetre range, given the ages just
derived for M4, M5 and M6.  The measurements reported in
Table~\ref{table measurements} and performed by \citet{Sandqvist95}
indicate negative indexes revealing, as expected for young clusters of
the derived ages, the presence of a non-thermal component. The
measurements also show that the indexes become increasingly positive
with increasing wavelength ($\alpha^{\rm 20cm}_{\rm 6cm} > \alpha^{\rm
  6cm}_{\rm 2cm}$). This is the signature of optically thick radio
emission. Indeed, it is known that UDHII regions can be optically
thick even at 6cm \citep{Kobulnicky99}.  In \citet{Sandqvist95}, even
though the 2\,cm fluxes are not given explicitly, one can extrapolate
their values using the 6cm flux and the 6cm/2cm spectral indexes
(given in their Table 2). The 2cm flux is interesting, since it is
both optically thin and likely to be free from a non-thermal
contribution.  Under this assumption, the measured radio fluxes
provide independent estimates of the ionising photon emission rates of
the clusters. From Table 2 from \citet{Sandqvist95}, we get the
following 2\,cm~fluxes for M4, M5 and M6: 1.75\,mJy, 0.61mJy,
1.70\,mJy respectively. The corresponding figures for the ionising
photon emission rates are 2.0\,$\times$10$^{52}$\,s$^{-1}$,
7.0\,$\times$10$^{51}$\,s$^{-1}$and 1.9\,$\times$10$^{52}$\,s$^{-1}$.
These values are slightly lower than, but in reasonable agreement with
those derived from the de-reddened emission lines.

\subsection{Final remarks about the derived YMC parameters}
\label{final age}

Along this procedure, the derivation of the cluster mass and the
derivation of the ionising photon emission rate do depend on the
validity of the assumption of a foreground uniform
extinction. Conversely, the age derivation is quite robust and relies
essentially on the disappearance of some emission lines relatively to
others. The factor to be matched is huge, one to two orders of
magnitude, and can in no way be understood in terms of measurement
uncertainties.

Would there be a possibility to explain the observed line ratios,
maintaining a very small age (around 1\,Myr)? Indeed, instead of the
Salpeter IMF which we consider, we might invoke an ad-hoc fancy IMF
with a low upper cutoff (at about 23\msol), but such a value is not
justified.  In addition, the observed negative spectral indexes of the
radio emission from the YMCs constitute an independent argument
favouring advanced ages, since the non-thermal component of the
centimetre emission traces the presence of supernovae, which only
occur after 3-4\,Myr. Therefore, we retain our age determinations as
providing the most sensible interpretation.

For the quoted age range for the clusters (6-8\,Myr), the
corresponding stellar masses for the three clusters are in the range
[0.6-3.]$\times10^7$\msol. These masses are derived from the
unextincted \Bg~fluxes and a Salpeter IMF with a lower boundary at
1\,\msol. This masses must be multiplied by 2.5 if the lower boundary
of the IMF is decreased to 0.1\,Myr.

The uncertainty on the estimate of the deredenned line luminosities ,
which we recall is performed under the assumption of foreground
extinction, is of the order of only 20$\%$, hence small with respect to
the uncertainty due to the uncertainties on the age and the IMF shape. 

We hence conclude that the clusters are 6-8\,Myr old, with masses of
the order $10^7$\msol~and ionising photon emission rates of several
$10^{52}$ s$^{-1}$.

All these elements concur to the stunning result that the three
MIR/radio sources in NGC1365 are among the most massive clusters
observed so far.  Surprisingly for their relatively ``advanced'' age,
their MIR emission indicates that they have not swept away all of
their surrounding material yet. Usually, clusters are believed to
remove very quickly the material in which they formed, within a few
million years. The peculiar case of these YMCs opens the question of
whether their extreme mass and their location close to the dust lane
of their host galaxy play a role in the fact that, at 7\,Myrs, they
still contain important amounts of gas.

\section{Fitting the YMC NIR/MIR SEDs}
\label{SED}

\begin{figure}[htbp]
\begin{center}
\includegraphics[width=9cm]{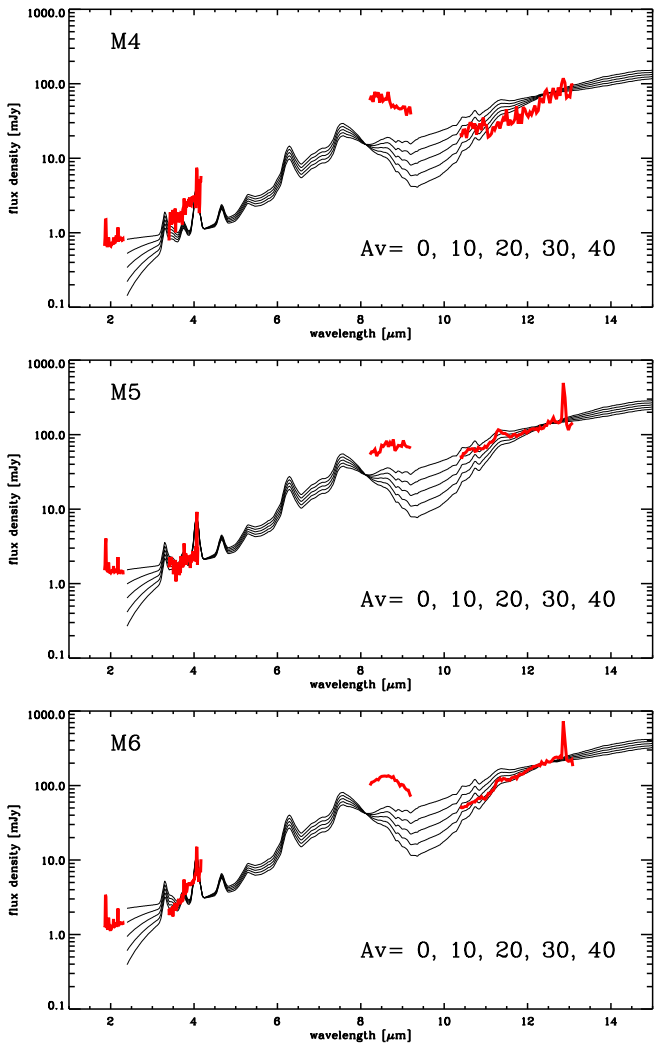}
\caption{Comparison between the observed spectra of NGC1365/M4, M5 and
  M6 to the ISO CVF spectra of R136 suffering from distinct foreground
  extinctions, from Av=0 to Av=40. The CVF spectra are normalised to
  the VISIR spectra at 12.5\micro.}
\label{comparison spectra}
\end{center}
\end{figure} 
\begin{figure*}[tbp]
\begin{center}
\includegraphics[width=12cm]{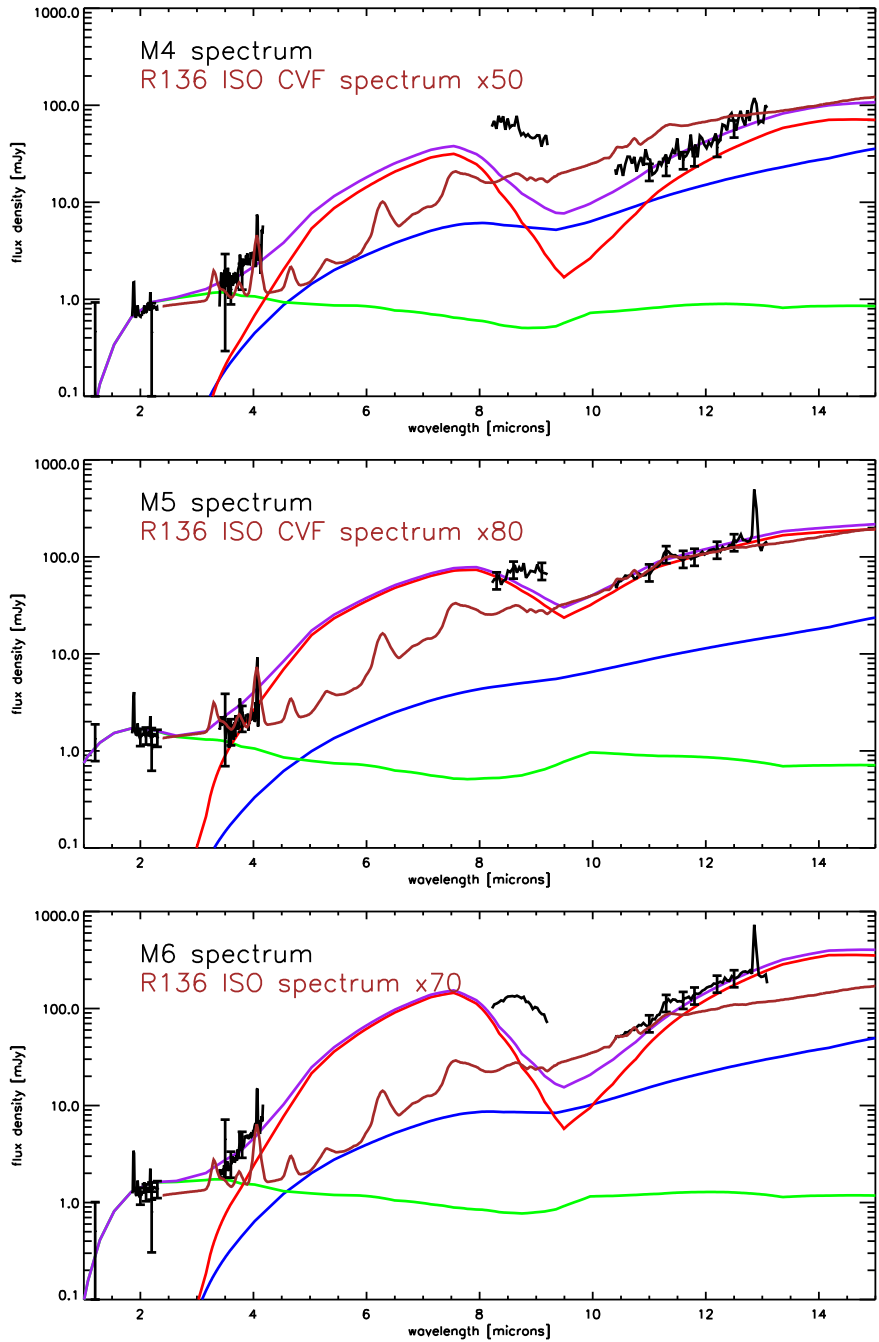}
\caption{Comparison of the observed spectra of M4, M5 and M6 (black)
  to the scaled ISO CVF spectrum of R136 (brown) and to the predicted
  spectra of the 2-component model described in the text. For the
  modelled spectra, the colour coding is the following: purple is the
  total spectrum (properly scaled for comparison with the source spectra); red
  is the total spectrum of the molecular cloud component (the
  optically thick one) component; blue is the spectrum of the dust (essentially very small grains, VSG)
  emission of the cirrus component (the optically thin one) and green
  is the spectrum of the[stellar+gas] component of the cirrus
  component}
\label{GRASIL models}
\end{center}
\end{figure*} 
\begin{figure}[htbp]
\begin{center}
\includegraphics[width=9cm]{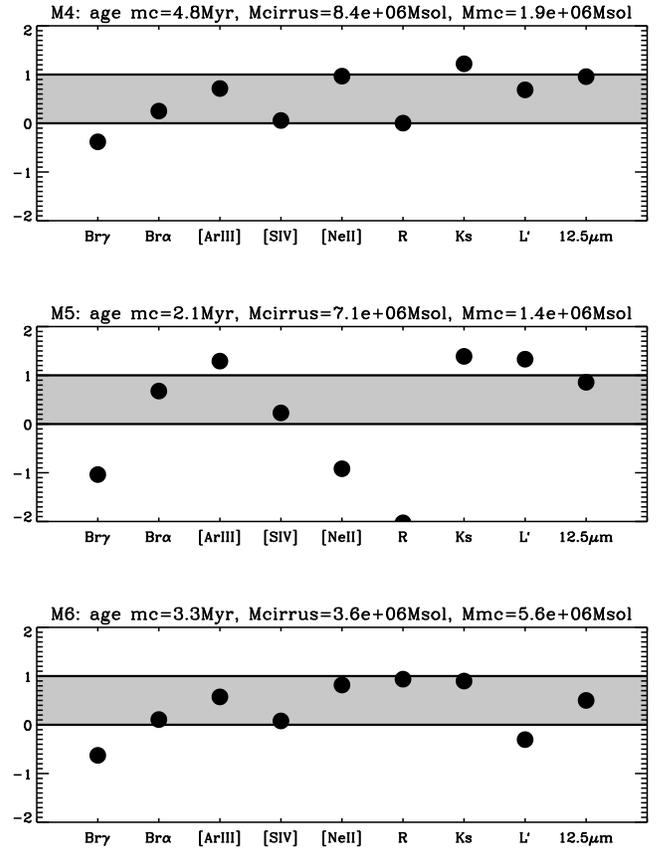}
\caption{Quality of the 9 predicted observables for the 2-component
  model for M4, M5 and M6 (see Sec. \ref{SED}). The observed
  uncertainty is shown with the grey stripe. If $obs_{min}$ and
  $obs_{max}$ represent the lower and upper limits of the error bar,
  then the value on the Y-axis is
  ($mod$-$obs_{min}$)/($obs_{max}$-$obs_{min}$), where $mod$ is the
  predicted value. For the observables for which we only measure an
  upper limit, then a value Y=0 is given to the corresponding point if
  the mode led value is lower than this limit.}
\label{fit result}
\end{center}
\end{figure} 
\begin{figure}[htbp]
\begin{center}
\includegraphics[width=9cm]{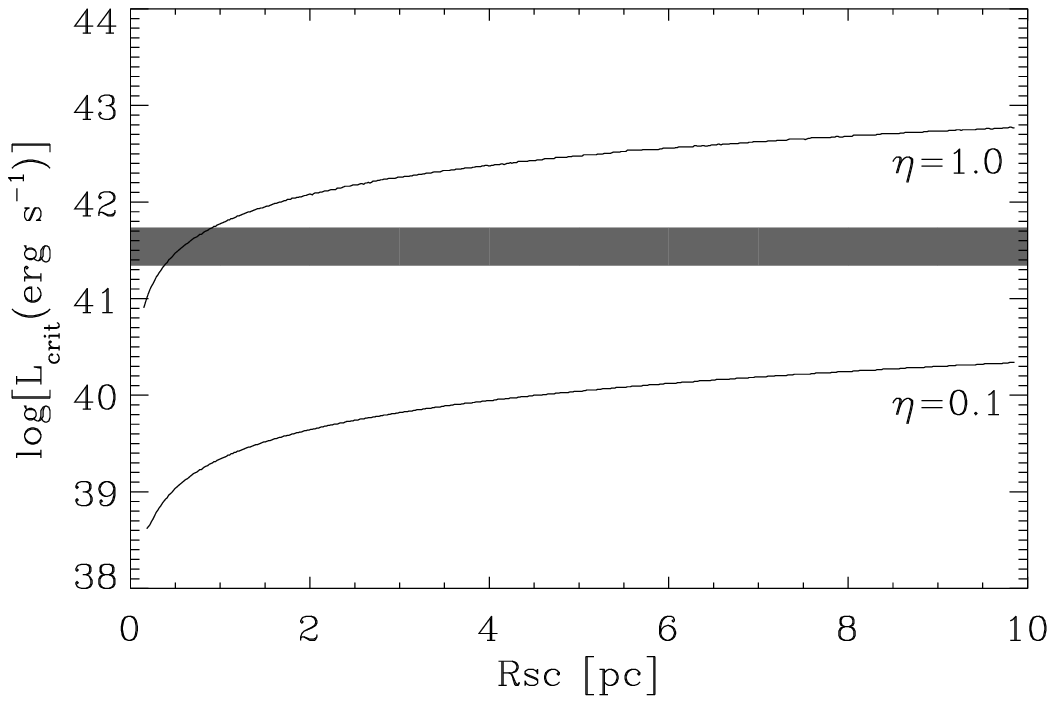}
\caption{Threshold mechanical luminosity for two heating efficiency
  \citep[reproduction of Fig. 4,b of][]{Wunsch07}. The horizontal line
  shows that, according to our estimates, the three MIR/radio clusters
  of NGC1365 lie in the bimodal regime region.}
\label{bi-modal}
\end{center}
\end{figure}

In this section, we discuss the fitting of the NIR/MIR SED of the YMCs. We later add the constraints from their emission line spectrum.

\subsection{Analysis of the NIR/MIR SEDs through comparison with R136}

First, we examine in Fig.~\ref{comparison spectra} a comparison of the
SEDs of M4, M5 and M6 with the ISO CVF spectrum of R136, scaled to the
distance of NGC1365, normalised at 12.5\micro, and subject to some
foreground extinction: we use a set of foreground extinctions in the
range Av=0 to Av=40.

One sees that in the 8-9\micro~window, the three NGC1365 YMCs are
brighter than R136 by factors between 2 and 10.  This factor cannot be
just explained by a calibration error for the 8-9\micro~range (See
Sec.~\ref{Measurements}) since, in the worst case, we evaluated a flux
uncertainty by a factor 2 for this region of the spectrum.  Even
though a full N-band spectrum would be mandatory to decisively settle
this difference, let us put forward the possibility that the high
8-9\micro~continuum is due to the presence in the YMCs of a bright
8.6\micro~PAH feature. In the usual PAH templates \citep[as used in
  GRASIL, ][]{Silva98}, the 12.7\micro~feature is much brighter than
the 8.6\micro~feature. However, the 8.6\micro~feature can be
significantly boosted for a high degree of ionisation of the PAH
molecules. Indeed, for ionised PAHs, the ratio of cross sections
$\sigma_{8.6}/\sigma_{12.7}$ is enhanced by a factor 7-8 with respect
to the neutral case \citep{Li01,Draine07}.

About the 10-13\micro~part of the SED, its slope can be reasonably
matched by a foreground extincted version of the R136 spectrum. In
particular the 10\micro-13\micro~VISIR spectra are well reproduced for
M5 and M6 (including the 11.3\micro~PAH feature). For M4, the
resemblance is not so obvious, but the M4 SED is also noisier.

In summary, the SED of M5 resembles that of R136, except in the
8\micro-9\micro~region where ionised PAH, boosting the
8.6\micro~feature, could be an explanation.  Therefore, M5 could be
seen as a scaled-up version of R136, from the SED point of
view. However, the emission line ratios discussed earlier clearly show
that M5 cannot have the same age as R136, but is at a more advanced
stage 6-8\,Myrs.  This is not a contradiction because the SED matching
is not an age estimator. The key issue is to understand how M5, at
~7\,Myrs, could have retained some dust component?

What is the conclusion of the comparison with R136 for M4 and M6? A
very distinctive feature observed in M4 and M6 is their steep
continuum rise between 3\micro~and 4\micro~, in contrary to the
spectrum of R136. This type of feature is reminiscent of the emission
from an optically thick dust component and suggests its presence in
the YMCs. Therefore, we investigate hereafter such a possibility,
using a more complex modelling tool.

\subsection{Modelling of the NIR/MIR SEDs}

Let us compare the SEDs we have in hand for the three YMCs to the SED
prediction of a cluster including two components: (a) an optically
thick component, which we will call the ``molecular cloud'' component,
with stellar mass M$_{mc}$ [which we would assume to be associated
  with the ``youngest'' stellar population in the cluster, in case the
  star formation would not have been instantaneous] and (b) an
optically thin component, which we will call the ``cirrus'' component,
with a stellar mass M$_{cirrus}$ [which we assume to be associated
  with the gas and with the ``oldest'' stellar population in the
  cluster].

We use \texttt{GRASIL} \citep{Silva98} to model the SED cluster.  In a
broad framework, \texttt{GRASIL} is designed to model the
spectro-photometric evolution of dusty galaxies, including a detailed
treatment of the radiative transfer through dust. It can be used to
simulate the evolution of the properties and emission of a stellar
population with the two components as described above: (a) the
``molecular cloud component'' for which the stars are still located at
the centre of their parent molecular cloud and for which a large
optical thickness is expected, and (b) the ``cirrus component'' in
which both the stars and the dusty gas show an extended distribution,
and for which the optical thickness is globally low.

Fig.~\ref{GRASIL models} illustrates how such a two-component model
can match the global features of the observed SEDs of the three
YMCs. This figure displays the observed SEDs for M4, M5 and M6 (in
black). For each source, the R136 spectrum scaled at around 2\micro~is
superimposed (in brown). As well, the different components of the
model are superimposed and shown under the following colour code: in
green is the joint stellar emission and continuum emission from the
HII region arising from the ``cirrus'' component, in blue is the dust
emission from the ``cirrus'' component, in red is the net emission
from the thick ``molecular cloud'' component (dust and stellar
emission, but essentially dominated by the dust) and in purple the
total emission.  The key parameter for the modelled ``molecular
cloud'' component is its optical depth. For each YMC, this optical
depth has been chosen in order to fit the slope in the red wing of the
silicate feature around 10\micro, observed in the YMC, and has been
derived after full treatment of the radiation transfer. The extinction
values Av which are required to reproduce the observed silicate
feature red wing slopes are very large: respectively 100, 70 and 130,
for M4, M5 and M6. Indeed, for the modelled ``molecular cloud''
component, dust is simultaneously the source of emission and
extinction. For details about the radiative transfer in
\texttt{GRASIL}, see \citet{Silva98}. Notice that the effect of such
an extinction cannot be compared in a straightforward manner to the
effect of the simple foreground extinction shown in
Fig.~\ref{comparison spectra} at the beginning of this section, and
estimated by comparing the YMC SED to the R136 spectrum.

Following this modelling, the NIR 3-4\micro~slope and the MIR slope
between 10\micro~and 13\micro~are both steep as a result of the
emission of the ``molecular cloud'' component and an important bump on
the SED in the 4-8\micro~range is predicted. This is consistent with
the observed rise of the SED in the 3-4\micro~range, in M4 and M6.

In conclusion, the NIR/MIR modelled SEDs in Fig.~\ref{GRASIL models}
provide qualitative matches for the observed SEDs in M4 and M6. In the
case of M5, the slope of the continuum in the NIR 3-4\micro~range is
overestimated in the model and the presence of a ``molecular cloud''
component does not seem to be mandatory.

Going one step further, let us investigate the possibility that the
``molecular cloud'' component represents a second stellar generation
(hence younger), still deeply embedded in dust.  For this purpose, we
perform fits of the two-component model predictions using the
following observational constraints: the \Bg, \Ba, [ArIII], [SIV] and
[NeII] fluxes, and the R, Ks, L' and 12.5\micro~flux densities. The
age of the cirrus component is set to 7\,Myr (derived earlier from the
analysis of the line ratios) and the optical thickness of the
``molecular cloud'' component to the values given above and derived
from the silicate red wing slope. The free parameters in the fits are
the age and mass M$_{mc}$ of the ``molecular cloud'' component (the
age is forced to be $\le6$\,Myr), the mass M$_{cirrus}$ of the
``cirrus'' component and the foreground extinction. Given the
observational uncertainties and because the centre of the error-bar
does not necessarily represent the most probable value of the
observable, different solutions can be found. Yet, the results are
globally identical. The quality of the fit for one solution is shown
graphically in Fig.~\ref{fit result}, and in fact, Fig.~\ref{GRASIL
  models} show this same solution. For each observable, the error-bar
has been scaled to the range [0-1]: the values of the modelled
observables are given on this same scale. A dot inside the grey zone
means that the modelled observable falls inside the error-bar.

The solutions shown in Fig.~\ref{fit result} are good for M4 and M6,
and marginally acceptable for M5. Indeed, Fig.~\ref{GRASIL models} was
already showing that the two-component model predicts a steep slope in
the 3-4\micro~range which is observed for M4 and M6, but not for
M5. For the fits presented in Fig.~\ref{fit result}, the age of the
``molecular component'', the stellar mass of the ``cirrus'' component
and the stellar mass of the ``molecular cloud'' component are
indicated above each plot.  For M4, the stellar mass in the
``molecular component'' is four time lower than in the ``cirrus''
component. On the contrary, it is almost two times greater in the case
of M6. Indeed, in Fig.~\ref{GRASIL models} the relative importance of the
``molecular component'' (red) to the ``cirrus component'' (green and
blue) is clearly greater in M6 than in M4.

The two-component model presented here, even though simplifying a
complex real situation and at the same time involving a large number
of parameters, allows an interesting qualitative match to the
observables. This is true to a good extent for M4 and M6. Notice that
the two-component model predicts a bump of the SED in the
4-8\micro~range, differing from the R136-type SED.  The SED of M5, on
the contrary, looks more like that of R136, and does not require the
presence of a thick ``molecular cloud'' component.  This might be
related to the fact that M5 is seen lying slightly outside the dust
lane, while M4 and M6 project right in the dust lane.  Indeed, the
fact that the pressure of the surrounding interstellar medium is
certainly higher around M4 and M6 than around M5, may have affected
their evolution and hence their intrinsic properties, and in peculiar
their ability to remove the material from which they formed initially.

\subsection{Conclusions on the YMCs parameters}
Let us summarise the conclusions reached after the successive steps of
the data analysis.

First, we have found that the three YMCs in NGC1365 are
compact clusters.  This is a solid conclusion, coming from a
direct comparison with the cluster R136 in the LMC.

Second, the age estimates of the YMCs come from three independent
channels:
 
(a) their NIR emission line ratios, with intense [NeII] and
undetectable [ArIII] and [SIV] lines, can be understood only if their
age is in the range 6-8 Myrs,

(b) their radio indexes are steep, suggesting that the non-thermal
contribution (supernovae remnants) is larger than the thermal
contribution (HII region): although it is not possible to extract a
precise value of the cluster age from its radio index, we know that
the clusters must be older than 3 Myrs,

(c) the detection of the CO absorption lines at 2.3\micro~in their
spectra tells us that the YMCs are older than 6 Myrs.

Therefore, we find the YMC ages to be around 7 Myrs. This
again is a fairly solid conclusion.

Third, given their ages and their ionising photon emission rates
(hence stellar luminosities), the stellar masses of the YMCs
  are found to be of the order of $10^7$\msol, implying a mass of
the initial molecular material from which they have been formed of at
least several $10^7$\msol. The molecular mass deduced from CO
observations by \citet{Sakamoto07}, of the order of 10$^9$\msol, is
indeed consistent with the figure we have obtained.


Fourth, we find that, after 6 to 8 Myr of evolution, still important
amounts of gaseous/dusty material, traced by the radio/MIR emission
and nebular lines, are found within or in the vicinity of the star
cluster. Since star clusters usually sweep away their gas on a much
shorter time scale, our observations present a case for trapping of
material in star clusters, at very high masses and in a dense
environment. This gas trapping can lead to peculiar evolution of the
star clusters by allowing subsequent star formation events to happen.



In the following, we discuss, in the framework of the theoretical
model by Tenorio-Tagle and co-workers, how such a gas trapping can be
explained.

%
%
%
%

\section{Gas trapping in the extremely massive clusters in NGC1365}      
\label{specific evolution}

Theoretical pieces of work by the group of Silich and Tenorio-Tagle
\citep{Tenorio07,Wunsch07,Silich07} analyse in details the
hydrodynamics of the matter re-inserted in young clusters by stellar
winds and supernovae ejecta. They show that in the case of massive and
compact clusters, the re-inserted material is exposed to strong
radiative cooling in the central parts of the clusters, because of the
high density of the gas in these regions. This strongly affects the
dynamics of the gas and can lead to a bi-modal hydrodynamic solution:
(i)) the matter injected inside a certain radius called the
\textit{stagnation radius} is accumulated and eventually becomes
gravitationally unstable leading to further star formation and, (ii))
outside the \textit {stagnation radius}, the re-inserted material
flows out of the cluster, building up a stationary wind
\citep{Wunsch07}.

One candidate for such a bi-modal gas behaviour has been identified in
M82 by \citet{Silich07}, using observations from \citet{Smith06}. The
observed parameters of the cluster, called M82-A1, imply such a
bi-modal hydrodynamic solution. The adopted model leads to a much
reduced rate of mass deposition in the interstellar matter, and a much
reduced wind terminal velocity, compared to the adiabatic wind model.

The three clusters NGC1365/M4, M5 and M6 are also excellent candidates
for displaying such a bi-modal behaviour: they are both very massive
and compact, hence the gas density in their central regions is
high. Also, the fact that with an age 7\,Myr, these clusters have
still retained an important amount of gas and dust, as shown by their
MIR emission, suggests that the gas removal mechanism has not been
efficient. Finally, the suspected ``molecular cloud'' component which
we derive from the SED modelling of M4 and M6 -- the two YMCs which
project onto the galaxy dust lane -- could trace on-going star forming
events, which recalls a prediction from the bi-modal
model. Nevertheless, in the case of M6, we find that $\rm M_{mc} >
M_{cirrus}$. This is incompatible with the idea that, for M6, the
``molecular component'' traces an ongoing star formation involving
only the material re-inserted by the ``first'' generation stars. The
higher pressure of the dust lane gas in which these two clusters seem
to be embedded could also have increased the efficiency of the
recycling of the matter into new generations of stars.

In order to investigate quantitatively whether the MIR/radio emitting
clusters in NGC1365 are viable candidates for this special regime, we
have used the results published in \citet{Wunsch07}. Simple analytic
formulae are provided in order to test if a cluster is undergoing such
a bi-modal hydrodynamic solution, or if it only drives a stationary
wind. Fig. 4 in \citet{Wunsch07} displays the threshold mechanical
luminosity as a function of the star cluster radius, that separates
single stationary wind solution from bi-modal solutions. The threshold
is shown to depend on the heating efficiency (parameter $\eta$~in
their work). This parameter represents the efficiency of the
thermalization of the gas.

To estimate the mechanical luminosity of the clusters, we use the
Starburst99 models \citep{Leitherer99}: for a Salpeter
$10^7$\msol~model, with solar metallicity, for an age range between 5
and 9\,Myr, the range of mechanical luminosities is
$10^{41.3}$~to~$10^{42.7}$\power.  The radii of the cluster stellar
component are lower than 10\,pc. This radius is directly measured for
M5, and assumed for M4 and M6, where no clear visible counterpart is
detected. In the radius vs. luminosity diagram of \citet{Wunsch07},
reproduced in Fig.~\ref{bi-modal}, the clusters clearly fall in the
bi-modal solution region.

The picture drawn here of the three bright MIR/radio sources seems
coherent.  In the massive reservoir of matter surrounding the nucleus
of NGC1365, and shown by the prominent dust lane, very massive and
compact clusters have formed. The hydrodynamics of the gas inside
these clusters (re-inserted and/or unused for star formation) is
susceptible to follow the bi-modal hydrodynamic solution described
above. This leads to a much less efficient removal of the gas, which
stagnates in the inner regions of the clusters, while it is ejected in
the outer regions with reduced velocities, compared to common
adiabatic cluster outflows. This could be traced by the velocity
gradient shown in Fig.~\ref{fig PV}, but we leave this analysis for a
future publication.

\section{Conclusion}      
\label{conclusion}

Let us summarise the main conclusions we have reached along
this work. 

The three MIR/radio sources in the circumnuclear star forming ring in
NGC1365 have been found to be very compact and very bright embedded
clusters. Under the simple assumption of a foreground extinction,
their MIR emission line spectrum points towards ages of about 7\,Myrs.
Such an age is also consistent with other age indicators like the
slope of their radio centimetre emission or the occurrence of thir NIR
spectrum of the CO absorption feature.  Given their ionising photon
emission rate and age, these YMCs are found to be extremely massive
objects, with a mass of at least 10$^7$\,\msol. They are the most massive
star clusters ever found so far. They must have formed from molecular
clouds of several 10$^7$\,\msol. This figure is consistent with the
presence within the NGC1365 central 2\,kpc diameter region, where they
are found, of large amounts of molecular gas, of the order of
10$^9$\msol, as derived from the CO molecule emission.
   

The remarkable output of our analysis of the data is the fact that
such embedded YMCs have retained an important amount of gas and dust,
in spite of their relatively advanced age, around 7\,Myrs. This a
quite puzzling and interesting fact, which may be related to the
extreme mass of these three YMCS. Indeed, some theoretical works have
analysed the fate of re-injected gas in very massive clusters and
found that the gas may be trapped and even lead to secondary star
forming events. The three MIR/radio clusters in NGC1365 are good
candidates to test the predictions of such models.



\begin{acknowledgements}

  We thank the daytime and nighttime support staff at Cerro Paranal
  Observatory, who made these observations possible, and the anonymous
  referee for her/his useful comments. We also warmly thank S. Silich
  and R. Wunsch for their comments on the manuscript. EG thanks the
  ESO fellowship program and the PCI program of ON/MCT (DTI/CNPq grant
  number 383076/07-2).

\end{acknowledgements}

\bibliographystyle{aa}

\end{document}